\documentclass[11pt, letter]{article}

\usepackage{comment}
\usepackage{pgf,pgfarrows,pgfautomata,pgfheaps,pgfnodes,pgfshade,multibox}
\usepackage{graphicx}
\usepackage{epstopdf}
\usepackage{amsmath}
\usepackage{amssymb}
\usepackage{amsfonts}
\usepackage{amsthm}
\usepackage{mathrsfs}
\usepackage{ccaption}

\definecolor{AV}{rgb}{0.65,0.0,0}
\definecolor{Ella}{rgb}{0.3,0,0.65}
\definecolor{Pau}{rgb}{0,0.5,0}
\definecolor{Jorge}{rgb}{0.2,0.7,0.7}

\usepackage[
   colorlinks=false, linkcolor=blue, urlcolor=blue,  filecolor=blue, citecolor=red, pdfstartview=FitV, pdftitle={}, pdfauthor={Pau Figueras}, pdfsubject={}, pdfkeywords={},pdfpagemode=None,bookmarksopen=true]{hyperref}

\usepackage{graphicx}
\usepackage{epstopdf}
\usepackage{epsfig,amssymb}

\newcommand{\rom}[1]{\mathrm{#1}}

\newcommand{\beq}{\begin{equation}}
\newcommand{\eeq}{\end{equation}}
\newcommand{\be}{\begin{equation}}
\newcommand{\ee}{\end{equation}}
\newcommand{\beqa}{\begin{eqnarray}}
\newcommand{\eeqa}{\end{eqnarray}}
\newcommand{\beqar}{\begin{eqnarray*}}
\newcommand{\eeqar}{\end{eqnarray*}}
\newcommand{\bea}{\begin{eqnarray}}
\newcommand{\eea}{\end{eqnarray}}

\newcommand{\la}{\lambda}

\newcommand{\al}{\alpha}

\newcommand{\Sg}{\Sigma}
\newcommand{\tht}{\theta}

\newcommand{\nn}{\nonumber}
\def\cV{\mathcal{V}}
\def \RR{{\mathbb{R}}}

\def\CA{{\cal A}}

\def\cA{\mathcal{A}}

\def\cF{\mathcal{F}}

\def\cL{\mathcal{L}}
\def\cM{\mathcal{M}}

\def\cP{\mathcal{P}}
\def\cQ{\mathcal{Q}}

\def\cV{\mathcal{V}}

\def\f{\frac}
\def\eps{\epsilon}
\def\beq{\begin{eqnarray}}
\def\eeq{\end{eqnarray}}
\def\mf{\mathfrak}
\def\Aone{\cA_{(1)}^1}
\def\Atwo{\cA_{(1)}^2}
\def\Azero{\chi_1}
\def\Fone{\cF_{(2)}^1}
\def\Ftwo{\cF_{(2)}^2}
\def\Fzero{\cF_{(1)}}
\def\axone{\chi_2}
\def\axtwo{\chi_3}
\def\fone{F_{(1)}^1}
\def\ftwo{F_{(1)}^2}
\def\gauge{A_{(1)}}
\def\strength{F_{(2)}}
\def\nn{\nonumber}
\def\half{\frac{1}{2}}
\def\no{\nonumber}

\makeatletter
\@addtoreset{equation}{section}

\begin{document}

\begin{titlepage}
\begin{flushright}
ArXiv: 0912.3199 [hep-th]\\
DCPT-09/87 \\
ULB-TH/09-43
\end{flushright}
\vskip 1cm
\begin{center}
{\Large Integrability of Five Dimensional Minimal Supergravity and Charged Rotating Black Holes}
\end{center}


\vspace{0.2cm}

\begin{center}
{\bf \large Pau Figueras$^\blacklozenge$, Ella Jamsin$^\Diamond$, Jorge V. Rocha$^\sharp$, and Amitabh Virmani$^\Diamond$}

\vspace{4mm} { \normalsize
$^\blacklozenge$Centre for Particle Theory, Department of Mathematical Sciences \\  University of Durham, South Road, Durham, DH1 3LE, UK\\

\vspace*{0.5cm}

\texttt{pau.figueras@durham.ac.uk} }

\vspace{4mm} { \normalsize
$^\Diamond$Physique Th\'eorique et Math\'ematique \\  Universit\'e Libre de
    Bruxelles\\and\\ International Solvay Institutes\\ Campus
    Plaine C.P. 231, B-1050 Bruxelles,  Belgium \\

\vspace*{0.5cm}

\texttt{ejamsin, avirmani@ulb.ac.be} }

\vspace{4mm} { \normalsize
$^\sharp$Centro Multidisciplinar de Astrof\'isica - CENTRA,\\
		Dept. de F\'isica, Instituto Superior T\'ecnico,\\
		Av. Rovisco Pais 1, 1049-001 Lisboa, Portugal \\

\vspace*{0.5cm}

\texttt{jorge.v.rocha@ist.utl.pt} }

\end{center}
\vspace{0.2cm}

\begin{abstract}
We explore the integrability of five-dimensional minimal supergravity in the presence of three
commuting Killing vectors.  We argue that to see the integrability structure of the theory one necessarily has to perform an Ehlers reduction to two dimensions. A direct dimensional reduction to two dimensions does not allow us to see the integrability of the theory in an easy way. This situation is in contrast with vacuum five-dimensional gravity.  We derive the Belinski-Zakharov (BZ) Lax pair for minimal supergravity based on a symmetric $7 \times 7 $ coset representative matrix for the coset $G_{2(2)}/(SL(2, \RR)\times SL(2, \RR))$. We elucidate the relationship between our BZ Lax pair and the group theoretic Lax pair previously known in the literature.  The BZ Lax pair allows us to generalize the well-known BZ dressing method to five-dimensional minimal supergravity. We show that the action of the three-dimensional hidden symmetry transformations on the BZ dressing method is  simply the group action on the BZ vectors. As an illustration of our formalism, we obtain the doubly spinning five-dimensional Myers-Perry black hole by applying solitonic transformations on the Schwarzschild black hole. We also derive the Cveti$\check{\rm c}$-Youm black hole by applying solitonic transformations on the Reissner-Nordstr\"om black hole.
\end{abstract}

\end{titlepage}

\vspace{0.5cm} \tableofcontents

\setlength{\unitlength}{1mm}

\section{Introduction}

The presence of $D-3$ commuting Killing symmetries allows us to reduce various $D$-dimensional gravity and supergravity theories into the form of three-dimensional non-linear sigma models coupled to three-dimensional gravity. The construction works for a large number of theories, ranging from the simplest case of four dimensional vacuum gravity giving rise to the $SL(2,\RR)/SO(2)$ coset model to eleven dimensional supergravity giving rise to an $E_{8(+8)}/SO(16)$ coset model. A classification of four dimensional theories leading to three-dimensional coset models was given in \cite{Breitenlohner:1987dg}. The higher dimensional origin of these coset models  was explored in \cite{Cremmer:1999du}.

Further reduction on another commuting Killing direction gives a two-dimensional dilaton gravity coupled to the corresponding non-linear sigma model now living in two dimensions \cite{Breitenlohner:1987dg}. The non-linear sigma model schematically represents the metric on the Killing directions of the higher dimensional spacetime and the matter sector of the higher dimensional theory.  This sigma structure plays a central role in the proofs of the celebrated black hole uniqueness theorems in four and five spacetime dimensions \cite{Robinson:1975bv,Mazur:1982db,Mazur:1983dc,Bunting:1983bu,Carter:1985bu,Hollands:2007aj,Hollands:2007qf,Tomizawa:2009ua,Amsel:2009et,Figueras:2009ci} (see also \cite{Chrusciel:2008js,Costa:2009hm}), and their recent generalisations to minimal supergravity in five dimensions \cite{Tomizawa:2009ua,Tomizawa:2009tb,Armas:2009dd}. It is well known in the literature \cite{BZ, Maison:1978es, Breitenlohner:1986um,  Nicolai:1991tt} (see \cite{Nicolai:1996pd} for a comprehensive review and further references) that these  two-dimensional coset models are classically completely integrable.
The underlying group theoretic structure is a crucial ingredient in establishing the integrability of the dimensionally reduced models.  Except in a few isolated cases, the integrability of these models has not yet been used as a solution generating technique. The notable exceptions are vacuum gravity  in various dimensions \cite{BZ} and four dimensional Einstein-Maxwell theory \cite{Alekseev}.

For five-dimensional vacuum gravity this line of investigation has led to an impressive progress in our understanding of stationary black holes with two rotational Killing vectors. It is expected that such an investigation for various supergravity theories will help us better understand the spectrum of black hole solutions in these theories.  Furthermore, it is likely that such an investigation will let us discover novel charged black hole solutions.  In this paper we take steps in this direction. Although our considerations are very general, and apply to any gravity theory that upon dimensional reduction gives rise to a coset model, for concreteness, we concentrate on the case of five-dimensional minimal supergravity.

A motivation behind focusing on five-dimensional minimal supergravity comes from the discovery of black rings  \cite{Emparan:2001wn},  and their subsequent supersymmetric  generalizations  \cite{Emparan:2004wy, Elvang:2004rt} (see \cite{Emparan:2006mm, Emparan:2008eg} for reviews and further references).   A non-supersymmetric non-extremal five parameter family of black ring solutions characterized by the mass, two angular momenta, electric charge, and dipole charge is conjectured to exist in  minimal supergravity \cite{Elvang:2004xi}.  At present, though, all known smooth black rings have no more than three independent parameters \cite{Elvang:2004rt, Elvang:2004xi}\footnote{In vacuum, the doubly spinning black ring solution of \cite{Pomeransky:2006bd} also has three independent parameters.}.
The solution of \cite{Elvang:2004xi} does not admit any smooth supersymmetric limit to the BPS black ring \cite{Elvang:2004rt}. It is likely that the integrability of minimal supergravity and the related inverse scattering technique explored in this paper will allow us to construct the most general black ring that will describe thermal excitations above the supersymmetric ring.

Another motivation for exploring integrability of supergravity theories comes from the fuzzball proposal. According to this proposal, a black hole geometry is a coarse grained description of its microstates. Some of theses microstates can be identified with smooth horizonless geometries with the same asymptotic charges as the black hole (see \cite{Mathur:2005zp} for reviews).
The fuzzball proposal has been mostly explored for BPS black holes, and large classes of smooth geometries corresponding to  microstates
 of certain supersymmetric  black holes
have been constructed. However, for most black holes a generic microstate need not admit a supergravity description \cite{deBoer:2009un}.

The situation for non-extremal black holes is much less developed. If the fuzzball proposal were to apply to non-extremal black holes as well, one would need to construct smooth horizonless non-BPS geometries with the same asymptotic charges as a non-extremal black hole. The construction of a large number of such geometries is a formidable task; at present, only a handful of such solutions are known \cite{Jejjala:2005yu, Giusto:2007tt, AlAlawi:2009qe, Bena:2009qv, Bobev:2009kn}. Although for supersymmetric black holes a detailed technology \cite{Gauntlett:2002nw} is available for constructing supersymmetric microstates, no systematic techniques are known to construct  non-supersymmetric microstates.
Having developed the inverse scattering method for supergravity theories, it is conceivable that one can
construct a large number of smooth horizonless non-BPS geometries. Such solutions  will not only advance our microscopic understanding of black holes, but also would shed light on how string theory can resolve the spacelike singularities of certain non-extremal black holes. Finally, we should also emphasize that the smooth horizonless geometries are of intrinsic interest from a purely gravitational point of view.

The inverse scattering method is a systematic procedure for generating new solutions from previously known solutions of a given set of non-linear equations.  The first step consists in finding a set of linear differential equations called the Lax pair whose integrability conditions are precisely the non-linear equations to be solved. If one then focuses on the special class of  solitonic solutions, the inverse scattering method provides an algebraic prescription to obtain analytically new solutions of the non-linear equations. For gravitational theories the solitonic solutions are the simplest and the most interesting ones.

The main results of this paper can be summarized as follows:
\begin{itemize}
\item We derive the Belinski-Zakharov (BZ) Lax pair for minimal supergravity based on a symmetric $7 \times 7$  coset representative matrix for the coset $G_{2(2)}/(SL(2,\RR) \times
SL(2,\RR))$. Our coset\footnote{Recall that $G_{2(2)}$ is the split real form of $G_2$. It is the only real form of $G_2$ that is relevant for our purposes. At the level of Lie algebras, following the standard notation, we denote the split real form of $\mf{g_2}$ as $\mf{g_{2(2)}}$.}  construction is largely based on the one used in \cite{G2}, but it differs in one important aspect, that the coset representative matrix $M$ is symmetric --- which is not the case in \cite{G2}. In \cite{G2} the
matrix $M$ is symmetric under generalized transposition, but not under the usual transpose. References
\cite{Bouchareb:2007ax, Clement2}
gave another coset construction for the coset $G_{2(2)}/(SL(2,\RR) \times
SL(2,\RR))$ where the matrix $M$ is also symmetric.  Our matrix $M$ shares several of the properties of the matrix $M$ of \cite{Bouchareb:2007ax, Clement2} though the two constructions are different. Refs \cite{Bouchareb:2007ax, Clement2} use different field variables and a different basis for the representation of $\mf{g_{2(2)}}$ than ours.
\item We elucidate the relationship between our BZ Lax pair and the group theoretic Lax pair previously known in the literature \cite{BZ, Maison:1978es, Breitenlohner:1986um,  Nicolai:1991tt, Nicolai:1996pd}.  We generalize the well-known BZ dressing method to five-dimensional minimal supergravity (modulo certain subtleties to be discussed in section \ref{stayinthecoset}).
\item We show that the action of the three-dimensional hidden symmetry transformations on the BZ dressing method is  simply the group action on the BZ vectors.
\item As an illustration of our formalism, we obtain the doubly spinning five-dimensional Myers-Perry black hole,   seen as a solution of minimal supergravity, by applying solitonic transformations on the Schwarzschild black hole. We also derive the Cveti$\check{\rm c}$-Youm black hole by applying solitonic transformations on the Reissner-Nordstr\"om black hole.
\item We argue that to see the integrability structure of the theory one necessarily has to perform an Ehlers reduction to two dimensions.  An Ehlers reduction is a two step reduction: first one reduces  the theory to three dimensions, dualizes all three-dimensional vectors into scalars, and then further reduces to two dimensions. A direct dimensional reduction to two dimensions does not allow us to see the integrability of the theory in an easy way. This situation is in  contrast with vacuum five-dimensional gravity.
\end{itemize}

The rest of the paper is organized as follows. In section \ref{sec:dimred} we present the dimensional reduction of five-dimensional minimal supergravity to two dimensions.   The integrability of this theory in explored in section \ref{sec:integrability}: in section \ref{inverse} we derive the BZ Lax pair; in section \ref{groupth} we show (following \cite{Breitenlohner:1986um}) that the BZ Lax pair is equivalent to the one used in \cite{Nicolai:1991tt, Breitenlohner:1986um}. In section \ref{sec:gen_BZ} we  generalize the BZ construction to minimal supergravity. The intuition behind how to use this construction is developed in section \ref{sec:factorspace}. Section \ref{sec:general_results} contains certain general results pertaining to the generalized BZ construction and the hidden $G_{2(2)}$ symmetry. In section \ref{sec:examples} we construct rotating black holes using the generalized BZ construction.  The argument that to see the integrability of five-dimensional minimal supergravity one needs to perform an Ehlers reduction is presented in section \ref{EvsMM}. We close with a discussion of open problems in section \ref{sec:open_problems}. Various technical details are relegated to appendices. In appendix \ref{CY} we present the Cveti$\check{\rm c}$-Youm solution in Weyl canonical coordinates.  We collect some general results for the Lie algebra $\mf{g_{2(2)}}$ in appendix \ref{genG2}.  In appendix \ref{rep} we give the representation of $\mf{g_{2(2)}}$ that we use and present a construction of a symmetric coset representative matrix $M$.  Finally, in appendix \ref{3form} we construct the three-form and the octonion structure constants preserved by our representation of $\mf{g_{2(2)}}$.

\section{Dimensional reduction to two dimensions}
\label{sec:dimred}

In this section we review the dimensional reduction of minimal five-dimensional supergravity from five to two dimensions. 
We assume the existence of three mutually commuting Killing vectors, $\xi_{a}$, $a=1,2,3$, so that $\mathcal{L}_{\xi_a}g_5=0$ and $\mathcal{L}_{\xi_a}F^5_{(2)}=0$ for all $a$, where $\mathcal{L}_{\xi_a}$ denotes the Lie derivative along $\xi_a$.
We first reduce the theory to three dimensions using two of these Killing vectors, dualize all three-dimensional one-forms into scalars, and then reduce the resulting three-dimensional theory to two dimensions on the remaining Killing vector.

\subsection{Reduction to three dimensions}
\label{3dreduction}

Reference \cite{Breitenlohner:1987dg} derived a list of three-dimensional symmetric space non-linear sigma-models obtained by dimensional reduction from a class of four dimensional gravity  theories coupled to abelian gauge fields and scalars\footnote{A detailed analysis of their group theoretical structure  was given in \cite{Breitenlohner:1998cv}, and the higher dimensional origin of these coset models  was explored in \cite{Cremmer:1999du}.}. In this paper we concentrate on the particular case of five-dimensional minimal (ungauged) supergravity, although we believe that our considerations can be readily generalized to other theories --- most likely to all theories considered in \cite{Breitenlohner:1987dg, Breitenlohner:1998cv, Cremmer:1999du}!

The dimensional reductions of five-dimensional minimal supergravity to three dimensions were first studied in \cite{Mizoguchi:1998wv,Cremmer:1999du}. When the reduction is performed over two spacelike Killing directions one obtains three-dimensional Lorentzian gravity coupled to the $G_{2(2)}/SO(4)$ coset model. On the other hand, when the reduction is performed over one timelike and one spacelike Killing direction one obtains three-dimensional Euclidean gravity coupled to the $G_{2(2)}/(SL(2,\RR)\times SL(2,\RR))$ coset model.  In this section, we briefly review these dimensional reductions. This presentation follows closely the one given in \cite{G2} but contains a little less details.

Five-dimensional minimal supergravity is the simplest supersymmetric extension of five-dimensional vacuum gravity. In the bosonic sector, it contains a metric $g_5$ and a gauge potential $A_{(1)}^5$ whose field strength is $F^5_{(2)}= dA^5_{(1)}$. The Lagrangian has the form of Einstein-Maxwell theory with a Chern-Simons term:
\begin{equation}
\cL_5 = R_5 \star 1 - \half\star  F^5_{(2)}\wedge F^5_{(2)}+\frac{1}{3\sqrt 3} F^5_{(2)}\wedge F^5_{(2)}\wedge A^5_{(1)} \, . \label{5dsugra}
\end{equation}

The reduction to three dimensions of the metric $g_5$ is performed using the following ansatz:
\begin{eqnarray}
ds^2_5 &=& e^{\frac{1}{\sqrt 3}\phi_1+\phi_2}ds^2_3 +\eps_2 e^{\frac{1}{\sqrt 3}\phi_1-\phi_2}(dz_4 + \Atwo)^2\nonumber\\
       & & +\eps_1 e^{-\frac{2}{\sqrt 3}\phi_1}(dz_5+\Azero dz_4+\Aone)^2\, ,
\label{metric5}
\end{eqnarray}
where the three-dimensional fields, namely the three-dimensional metric $g_3$, the two dilatons $\phi_1$ and $\phi_2$, the axion $\chi_1$, and the two Kaluza-Klein one-form potentials $\Aone$ and $\Atwo$, do not depend on $z_4$ and $z_5$ coordinates. One can also think of this reduction as a two step process. The first step being the reduction from five to four dimensions over $z_5$, and the second being the reduction from four to three dimensions over $z_4$. In each step, the reduction can be performed over either a spacelike or a timelike Killing direction.  The sign $\epsilon_{i}$ is $+1$ when the reduction is performed over a spacelike direction, and $-1$ for a timelike direction.  We denote the field strengths associated to  $\Azero$, $\Aone$, and $\Atwo$ by $\Fzero$, $\Fone$, and $\Ftwo$ respectively. They are defined to be,
\begin{eqnarray}
\Fzero&=& d\Azero\, , \nonumber\\
\Fone&=& d\Aone+ \Atwo\wedge d\Azero\, ,\nonumber \\
\Ftwo&=&d\Atwo \, .\nonumber
\end{eqnarray}

The reduction ansatz of the five-dimensional gauge potential $A^5_{(1)}$ is taken to be
\begin{eqnarray}
\label{pot5}
A^5_{(1)} = \gauge+ \axtwo dz_4+ \axone dz_5,
\end{eqnarray}
where similarly the three-dimensional gauge potential $\gauge$ and the two axions $\axone$ and $\axtwo$ are independent of $z_4$ and $z_5$. The associated field strengths $\strength$, $\fone$ and $\ftwo$ are defined to be,
\begin{eqnarray}
\fone&=& d\axone,\nn\\
\ftwo &=&d\axtwo-\Azero d\axone ,\\
\strength&=& d\gauge- d\axone\wedge (\Aone- \Azero \Atwo)- d\axtwo\wedge \Atwo.\nn
\end{eqnarray}

The next step in the reduction, in order to see the full hidden symmetry, is to define the axions  $\chi_4$, $\chi_5$, and $\chi_6$  dual to the one forms $\gauge$, $\Aone$, and $\Atwo$.
This can be done by introducing the dual one-form field strengths $G_{(1)4}$, $G_{(1)5}$ and $G_{(1)6}$ for the three axions:
\begin{eqnarray}
e^{-\vec\alpha_4 \cdot \vec\phi}\star \strength&\equiv &G_{(1)4} = d\chi_4+\frac{1}{\sqrt 3} (\chi_2 d\chi_3 - \chi_3 d\chi_2), \nonumber \\
\eps_1 e^{-\vec\alpha_5 \cdot \vec\phi} \star  \Fone& \equiv &G_{(1)5} = d\chi_5 - \chi_2 d\chi_4 + \frac{1}{3\sqrt{3}} \chi_2 (\chi_3 d\chi_2 - \chi_2 d\chi_3), \label{dualfields}\\
\eps_ 2 e^{-\vec\alpha_6 \cdot \vec\phi}\star  \Ftwo  &\equiv &G_{(1)6} = d\chi_6- \chi_1 d\chi_5+ (\chi_1\chi_2 - \chi_3)d\chi_4\nonumber\\
&& \qquad \quad \: + \:  \frac{1}{3\sqrt{3}} (-\chi_1\chi_2+\chi_3) (\chi_3 d\chi_2 - \chi_2 d\chi_3) \nonumber.
\end{eqnarray}
In terms of the new variables, $\phi_1$, $\phi_2$, $\chi_1,\dots,\chi_6$, the Lagrangian becomes
\begin{eqnarray}
\cL &=& R \star 1- \half \star d\vec\phi\wedge d\vec\phi - \half \eps_1\eps_2 e^{\vec\alpha_1 \cdot \vec\phi} \star  d\chi_1 \wedge d \chi_1 - \half  \eps_1 e^{\vec\alpha_2 \cdot \vec\phi}\star  d\chi_2\wedge d\chi_2\nonumber\\
&&- \half \eps_2 e^{\vec\alpha_3 \cdot \vec\phi}\star  (d\chi_3-\chi_1 d\chi_2)\wedge (d\chi_3-\chi_1 d\chi_2) + \half\eps_t e^{\vec\alpha_4 \cdot \vec\phi}\star G_{(1)4} \wedge  G_{(1)4} \nonumber\\
&&+  \half \eps_1 \eps_t e^{\vec\alpha_5 \cdot \vec\phi}\star  G_{(1)5} \wedge  G_{(1)5} +\half\eps_2 \eps_t e^{\vec\alpha_6 \cdot \vec\phi}\star  G_{(1)6} \wedge  G_{(1)6}~,\label{Lcoset_implicit}
\end{eqnarray}
where $\eps_t$ denotes the  signature of the three-dimensional metric. It appears in this expression because of the relation $\star \star \omega_{(1)} = \eps_t \omega_{(1)}$ for any one-form $\omega_{(1)}$. The six doublets  $\vec\alpha_1, \dots, \vec\alpha_6$ correspond precisely to the six positive roots of the exceptional Lie algebra $\mf{g_2}$, given in appendix \ref{genG2}. One can note from the Lagrangian (\ref{Lcoset_implicit}) that for each $i$ the axion $\chi_i$ is associated to the root $\alpha_i$.

To summarize, the three-dimensional theory is determined by a three-dimensional metric and a set of eight scalar fields: two dilatons $\phi_1$ and $\phi_2$ and six axions $\chi_1,\dots,\chi_6$.

\subsection*{The non-linear $\sigma$-model for $G_{2(2)} / \tilde{K} $}
\label{sigma}

It turns out that the Lagrangian (\ref{Lcoset_implicit}) can be rewritten as
\beq \cL = R\star 1 + \cL_{\rom{scalar}} \, , \eeq
where $\cL_{\rom{scalar}}$ is the Lagrangian of a non-linear $\sigma$-model for the coset $G_{2(2)}/\tilde{K}$, with an appropriate subgroup $\tilde{K}$
depending on the signature $\epsilon_{1,2}$ of the reduced dimensions.  We can write a coset representative $\cV$ for the coset $G_{2(2)}/\tilde{K}$ in the Borel gauge by exponentiating the Cartan and positive root generators of $\mf{g_{2(2)}}$ with the dilatons and axions as coefficients.  We can make contact with the reduced Lagrangian (\ref{Lcoset_implicit}) by choosing the coset representative to be \cite{G2}
\begin{equation}
\mathbb \cV = e^{\half \phi_1 h_1 + \half \phi_2 h_2} e^{\chi_1 e_1 }e^{-\chi_2 e_2 +\chi_3 e_3}e^{\chi_6 e_6} e^{\chi_4 e_4 -\chi_5 e_5},
\label{coset}
\end{equation}
where the representation of $\mf{g_{2(2)}}$ that we use is given in appendix \ref{rep}. For more group theoretic details on the construction and properties of the coset representative $\cV$ we refer the reader to \cite{G2}.
Next we define the matrix $M$ as
\be
\label{M}
M = S^{T} \cV^T \eta \cV S ~,
\ee
where $\eta$ and $S$ are constant matrices whose explicit expressions are given in appendix \ref{rep} for the choice $\epsilon_1 = -1,\epsilon_2 = +1$, to which we specialize from now on. (For other choices of $\epsilon$'s, the matrix $\eta$ needs to be adapted.)
The matrix $M$ is symmetric by construction. Under global $G_{2(2)}$ transformations it transforms as
  \be
  M \rightarrow M_g = (S^{-1}g S)^{T} M (S^{-1}g S)~ \quad \mbox{for} \quad g \in G_{2(2)}. \label{g2transformation}
  \ee
It can be easily checked by an explicit calculation that the scalar part of the reduced Lagrangian (\ref{Lcoset_implicit}) is also given by
\be
\cL_{\rom{scalar}} = - \frac{1}{8} \mbox{Tr}\left(\star (M^{-1}d M) \wedge (M^{-1} d M)\right).
\label{lagM}
\ee
We will see below that the matrix $M$ plays the role of the metric on the Killing fields in the standard BZ construction.  The vacuum truncation of the matrix $M$ (i.e., setting $\chi_2 = \chi_3 = \chi_4 = 0$) is block diagonal\footnote{The matrix $S$ above was chosen precisely to ensure that this is the case. This block diagonal form simply refers to the fact that the seven dimensional representation of $G_{2(2)}$ branches into $SL(3, \RR)$ representations as $\mathbf{7}= \mathbf{\bar{3}} + \mathbf{1} + \mathbf{3}$.}:
\beq
M  = \left(
\begin{array}{ccc}
M_{\rom{SL(3)}}^{-1} & 0 & 0 \\
0  & 1 & 0 \\
0 & 0 & M_{\rom{SL(3)}}
\end{array}
\right) \, ,
\label{vacuum_mt}
\eeq
where
\beq
M_{\rom{SL(3)}} = \left(
\begin{array}{ccc}
- \frac{1+e^{\sqrt{3} \phi _1+\phi _2} \chi _5{}^2}{e^{\frac{2 \phi _1}{\sqrt{3}}}} & - \frac{\chi _1+e^{\sqrt{3} \phi _1+\phi _2} \chi _5 \chi _6}{e^{\frac{2 \phi _1}{\sqrt{3}}}} & -e^{\frac{\phi _1}{\sqrt{3}}+\phi _2} \chi _5 \\
 - \frac{\chi _1+e^{\sqrt{3} \phi _1+\phi _2} \chi _5 \chi _6}{e^{\frac{2 \phi _1}{\sqrt{3}}} }  &  \frac{-e^{\phi _2} \chi _1{}^2+e^{\sqrt{3} \phi _1} \left(1-e^{2 \phi _2} \chi _6{}^2\right)}{e^{\frac{2 \phi _1}{\sqrt{3}}+\phi _2} } & -e^{\frac{\phi _1}{\sqrt{3}}+\phi _2} \chi _6 \\
 -e^{\frac{\phi _1}{\sqrt{3}}+\phi _2} \chi _5 & -e^{\frac{\phi _1}{\sqrt{3}}+\phi _2} \chi _6 & -e^{\frac{\phi _1}{\sqrt{3}}+\phi _2}
\end{array}
\right).
\label{chisl3}
\eeq
This $M_{\rom{SL(3)}}$ is identical to the matrix $\chi$ used in \cite{Giusto:2007fx}. Using a different representation of $\mf{g_{2(2)}}$, references \cite{Bouchareb:2007ax, Clement2} gave another coset construction for the coset $G_{2(2)}/(SL(2, \RR) \times SL(2, \RR))$ where the matrix $M$ is also symmetric and takes the form (\ref{vacuum_mt}) for vacuum gravity. Further properties of the matrix $M$ are given in appendix \ref{rep}.

In most references, and in particular in \cite{G2} where a similar discussion is presented, the matrix $M$ is rather defined as $\cV^{\sharp}\cV$, where $\sharp$ denotes the generalized transposition, which defines the subgroup on which the coset is taken. That expression is related to the one we use in this paper in an easy way. First, one can see that in our representation
\beq
\label{defsharp}
\cV^{\sharp}=\eta^{-1}\cV^{T}\eta.
\eeq
As a consequence, the two choices of definitions for $M$ are related in the following way, where we keep the notation $M$ for the expression (\ref{M})
\beq
\mathcal{M}:=\cV^{\sharp}\cV=\eta^{-1} \, (S^{T})^{-1} M S^{-1} ~.
\eeq

\subsection{From three to two dimensions}
\label{threetwo}

The reduction to two dimensions is performed by dropping all dependence on the third variable $z^3$ and using the usual ansatz for the metric
\beq
g_{\mu \nu}=\left(\begin{array}{cc}\xi^2 \bar g_{m n} + \rho^2 B_m B_n & \rho^2 B_m \\ \rho^2 B_n & \rho^2\end{array}\right)
\eeq
where $\mu,\nu$ are indices in three dimensions, $m, n$ are two-dimensional indices, and $\bar g_{mn}$ is the two-dimensional metric. Two comments are in order:
\begin{itemize}
\item The two-dimensional Kaluza-Klein vector $B_m$ can be dropped under some weak conditions  (see \cite{Wald:1984rg,Emparan:2001wk} in the context of vacuum gravity and \cite{Tomizawa:2009ua} in the context of five-dimensional minimal supergravity).   This implies that the two-dimensional spaces orthogonal to the commuting Killing vector fields are integrable submanifolds.
\item Any two-dimensional metric can be written locally as a conformal factor times the Minkowski metric. As a consequence, one can take $\bar g_{mn} = \delta_{mn} $ (recall that the three-dimensional space is Euclidean), while absorbing the conformal factor in $\xi$.
\end{itemize}
Thus, the ansatz simplifies to the diagonal one
\beq
g_{\mu\nu}=\left(\begin{array}{cc}\xi^2  \delta_{mn} & 0 \\0 & \rho^2\end{array}\right).
\label{gmunu}
\eeq
Taking moreover into account that $M^{-1} \partial_3 M=0$, the Lagrangian (\ref{lagM}) becomes \cite{Breitenlohner:1987dg}
\begin{equation}
\cL = 2\xi^{-1}\star d\rho\wedge d\xi-\frac18\rho\mathrm{Tr}(\star (M^{-1} dM)  \wedge (M^{-1} dM))
\end{equation}
where now the Hodge dual $\star$ and the differential $d$ are taken over the two-dimensional flat space.

The equation of motion for the field $\rho$ is given by
\beq
d\star d\rho&=&0\label{eom1}
\eeq
From (\ref{eom1}) one deduces that $\rho$ is a harmonic function; therefore, one can choose it to be one of the coordinates on the two-dimensional space, $z_1 = \rho$. We take the second coordinate $z_2$ to be $z_2=z$ such that $d z=-\star d\rho$.

In coordinates ($\rho, z$) the equations for the matrix field $M$ can be rewritten as
\beq
 d(\star\rho\, M^{-1}dM)=0 \,.\label{prebasiceom1}
\eeq
We find it useful to rewrite this equation explicitly in terms of the two-dimensional variables as
\begin{equation}
 \partial_{\rho}(\rho\,\partial_\rho M\,M^{-1})+\partial_{z}(\rho\,\partial_z M\,M^{-1})=0\,. \label{basiceom}
\end{equation} Defining the matrices $U$ and $V$ as follows,
\begin{equation}
 U=\rho\,(\partial_\rho M)\,M^{-1}\,,\qquad V=\rho\,(\partial_z M)\,M^{-1}\,,
\label{defUV}
\end{equation}
the equation of motion \eqref{basiceom} becomes,
\begin{equation}
 \partial_\rho U+\partial_z V=0\,.
 \label{UVeq}
\end{equation}
Finally, the equations for $\xi$ can be written as the following system \cite{Breitenlohner:1987dg}
\beq
\xi^{-1}\partial_\rho\xi=\frac1{16\rho}\mbox{Tr}(U^2-V^2), \qquad  \xi^{-1}\partial_z\xi=\frac1{8\rho}\mbox{Tr}(U V) .
\label{lambdaeq}
\eeq
The equations for $\xi$ satisfy the integrability condition $\partial_\rho \partial_z \xi = \partial_z \partial_\rho \xi$ as a consequence of (\ref{UVeq}). Therefore, once $M$ is known, the function $\xi$ is determined by a line integral, up to an integration constant. Moreover, while reducing the theory on one timelike and two rotational Killing vectors the $\rho$ and $z$ components of the gauge-field can be taken to be zero \cite{Tomizawa:2009ua}, \be A_\rho = A_z = 0~.\ee

Equations (\ref{UVeq}) and (\ref{lambdaeq}) are the equations for a completely integrable two-dimensional sigma model. We will comment on the group theoretic structure of this sigma model in some detail in the following sections.

\section{Integrability and the linear system}
\label{sec:integrability}

In this section we explore the integrability of the theory. In  section \ref{inverse} we derive the BZ Lax pair for five-dimensional minimal supergravity. This formulation of the Lax pair is equivalent to the one used in \cite{Nicolai:1991tt, Breitenlohner:1986um}, but it is better suited to generate new solutions using the inverse scattering method. In section \ref{groupth} we discuss the interrelation between the Lax pair introduced in section \ref{inverse} and that of \cite{Nicolai:1991tt, Breitenlohner:1986um}. We also briefly discuss the solution generating technique of  \cite{Nicolai:1991tt, Breitenlohner:1986um}.

\subsection{The Belinski-Zakharov approach}
\label{inverse}

The approach of Belinski and Zakharov \cite{BZ}, well known for vacuum gravity in various dimensions, can be readily generalized to minimal 5d supergravity using the equations deduced in section \ref{threetwo}. Focusing on  vacuum gravity in the presence of $D-2$ commuting Killing vector fields, one can show \cite{Wald:1984rg, Emparan:2001wk} that, under natural suitable conditions, the metric admits the form
\be
ds^2 = G_{\bar \mu \bar \nu} dx^{\bar \mu} dx^{\bar \nu} + e^{2\nu}(d\rho^2 + dz^2)~,
\label{Killing+conformal}
\ee
where $x^{\bar \mu} = z_3, z_4, z_5$.
Furthermore, without loss of generality we can choose coordinates so that
\be
\det G = - \rho^2.
\ee
Recall, from the previous section, that the coordinate $z$ is defined as the harmonic dual of $\rho$.\footnote{For vacuum spacetimes containing a black hole, it has been rigorously proven that these coordinates are globally defined in the domain of outer communications \cite{Chrusciel:2008js}. See \cite{Costa:2009hm} for a generalization to the four dimensional Einstein-Maxwell theory.}  Integrations of the three Killing directions yields a two-dimensional non-linear sigma model that is completely integrable.

The vacuum Einstein equations divide into two groups, one for the metric components along the Killing directions,
\be
\partial_\rho \tilde U + \partial_z \tilde V = 0,
\label{eomG}
\ee
with
\be
 \tilde U = \rho (\partial_\rho G ) G^{-1},  \qquad  \tilde V = \rho (\partial_z G) G^{-1}~,
\ee
and the second group of equations for $\nu$
\be
\partial_\rho \nu =- \frac{1}{2\rho} + \frac{1}{8\rho}\mbox{Tr}(\tilde U^2-\tilde V^2)~, \qquad
\partial_z \nu=\frac1{4\rho}\mbox{Tr}(\tilde U \tilde V)~.
\label{nueq}
\ee
These equations (\ref{eomG}-\ref{nueq}) are the starting point of the BZ construction. It is clear that they are formally identical to equations (\ref{defUV}-\ref{lambdaeq}) for 5d minimal supergravity\footnote{\label{footnotenu}Explicitly, equation~(\ref{nueq}) is converted into the form~(\ref{lambdaeq}) if one replaces $G$ by $M$ and $e^\nu$ by $\xi^2/\sqrt{\rho}$.}, the main difference being that in the latter case the matrix $M$ plays the role of the Killing part $G$ of the spacetime metric. Therefore, the Lax pair we are seeking can be immediately derived from the BZ Lax pair,
\begin{equation}
 D_1\Psi=\frac{\rho\,V-\lambda\,U}{\lambda^2+\rho^2}\,\Psi\,,\qquad
D_2\Psi=\frac{\rho\,U+\lambda\,V}{\lambda^2+\rho^2}\,\Psi\,,
\label{Laxpair}
\end{equation}
where $\lambda$ is the spacetime dependent spectral parameter, and $\Psi(\lambda,\rho,z)$ is the generating matrix such that the matrix
$M$ can be extracted from $\Psi$ as
\be
M(\rho, z)  =  \Psi(0, \rho, z)\, .
\label{genmat}
\ee
$D_1$ and $D_2$ are the commuting differential operators introduced by BZ:
\begin{equation}
D_1=\partial_z-\frac{2\,\lambda^2}{\lambda^2+\rho^2}\,\partial_\lambda\,,\qquad
D_2=\partial_\rho+\frac{2\,\lambda\,\rho}{\lambda^2+\rho^2}\,\partial_\lambda\,.
\label{defD1D2}
\end{equation}

Indeed, since $[D_1,D_2]=0$ the compatibility conditions of \eqref{Laxpair} are given by
\begin{equation}
 [D_1,D_2]\Psi=\frac{1}{\lambda^2+\rho^2}\Big\{\lambda(\partial_\rho U+\partial_z V)+V+\rho(\partial_z U+\partial_\rho V)+[U,V]\Big\}\Psi=0\,.
\end{equation}
We observe that the term in this expression proportional to $\lambda$ is just the equation of motion \eqref{basiceom}, while the remaining term corresponds to the integrability condition that follows from \eqref{defUV}. This guarantees that the solution of the system \eqref{Laxpair} yields a solution of the original (non-linear) equation \eqref{basiceom}.

An important feature when generalizing the BZ construction to 5d minimal supergravity is the step in the dimensional reduction that consists in dualizing the 3d vectors into scalars. Such a dimensional reduction is called an Ehlers reduction, while when the reduction is performed without dualization it is called Matzner-Misner. Without the dualization, one cannot see the integrability of 5d minimal supergravity in two dimensions. On the other hand, the dualization is not necessary for vacuum gravity, as well as for some other supergravity theories. These  subtleties can be better understood in terms of group theory and Dynkin diagrams and are discussed in section \ref{EvsMM}.

The upshot of this derivation is that the technology developed by BZ for constructing solitonic solutions of \eqref{Laxpair} can be applied to five-dimensional minimal supergravity (modulo certain subtleties to be discussed in section \ref{stayinthecoset}). In fact, most likely, the BZ construction can be applied to all theories considered in \cite{Cremmer:1999du}, which reduce to non-linear sigma models in three dimensions and whose equations of motion are of the form \eqref{basiceom} when reduced to two dimensions.

\subsection{The Breitenlohner-Maison approach}
\label{groupth}

A somewhat different approach for describing integrability of the two-dimensional coset models was taken in \cite{Breitenlohner:1986um, Nicolai:1991tt}. It has the advantage to make the underlying symmetries more transparent. Here we summarize salient aspects of this approach and show that their Lax pair is equivalent to the Lax pair given above. Our presentation follows \cite{Breitenlohner:1986um, Nicolai:1991tt}. See also \cite{Lu:2007zv, Schwarz:1995td}.

In this approach, in order to write the Lax pair, one first defines a new matrix $\hat \cV(\gamma,x)$ which depends on a spacetime dependent spectral parameter $\gamma$ and satisfies $\hat \cV(\gamma=0,x)=\cV(x)$. Here $x$ denotes collectively the coordinates on the two-dimensional flat base space. The Lax pair is now given by
\begin{equation}
d\hat \cV\,\hat \cV^{-1}=\cQ+\frac{1-\gamma^2}{1+\gamma^2}\,\cP-\frac{2\,\gamma}{1+\gamma^2}\,\star\cP\,,
\label{LaxV1}
\end{equation}
where
\begin{equation}
\cQ=\frac{1}{2}\big[d\cV\,\cV^{-1}+\tau(d\cV\,\cV^{-1})\big]\,,\qquad
\cP=\frac{1}{2}\big[d\cV\,\cV^{-1}-\tau(d\cV\,\cV^{-1})\big]\,,
\label{LaxV2}
\end{equation}
 $\tau$ is the involution of the Lie algebra $\mf{g_{2(2)}}$ associated to the coset $G_{2(2)}/(SL(2, \RR) \times SL(2, \RR))$, and $\star$ denotes the Hodge dual on the two-dimensional flat base space.

The integrability condition for these equations is equivalent to the equations of motion (\ref{prebasiceom1}), provided that the spectral parameter $\gamma$ obeys the differential equation
\begin{equation}
(1-\gamma^2)\,d\gamma+2\,\gamma\star d\gamma=\frac{\gamma(1+\gamma^2)}{\rho}\,d\rho\,.
\end{equation}
This equation can be easily solved and the solution is given by
\begin{equation}
\frac{1}{\gamma}-\gamma=\frac{2(w-z)}{\rho} \quad \Rightarrow \quad \gamma_{\pm}(w,x)=\frac{1}{\rho}\left[(z-w)\pm\sqrt{\rho^2+(z-w)^2}\right]\,,
\end{equation}
where $w$ is a constant, the so-called `spacetime \emph{independent} spectral parameter'. Defining the matrix $X(\gamma,x)=\cV(x)^{-1}\,\hat \cV(\gamma,x)$, we find that the Lax pair can be rewritten as
\begin{equation}
dX X^{-1}=-\frac{\gamma^2}{1+\gamma^2}\,\CA-\frac{\gamma}{1+\gamma^2}\,\star\CA\,, \label{LaxX}
\end{equation}
where $\CA=M^{-1}dM$. The Lax equations (\ref{LaxV1}) are the ones used in \cite{Breitenlohner:1986um, Nicolai:1991tt}.

\subsection{Equivalence between the Lax pairs}
\label{equivalence}

To show that these equations are equivalent to the BZ Lax pair, first notice that $\gamma_-=-\mu/\rho$ and $\gamma_+=-\bar\mu/\rho$, where
\begin{equation}
\mu=\sqrt{\rho^2+(z-w)^2}-(z-w)\,,\qquad  \bar\mu=-\rho^2/\mu\,. \label{solitons}
\end{equation}
These correspond to the `solitons' and `anti-solitions' in the context of the BZ construction for vacuum gravity. This motivates us to define \cite{Breitenlohner:1986um} a new spectral parameter
\be
\lambda \equiv -\rho\,\gamma~,
\ee so that now $\lambda$ corresponds to either a soliton $(\lambda_-=\mu)$ or an antisoliton $(\lambda_+=\bar\mu)$.  Notice that the new spectral parameter satisfies
\begin{equation}
d\lambda=\frac{2\,\lambda\,\rho}{\lambda^2+\rho^2}\left(d\rho-\frac{\lambda}{\rho}\,dz\right)\,. \label{eqlambda}
\end{equation}
We can rewrite the Lax pair for the matrix $X$, \eqref{LaxX}, in terms of the new spectral parameter $\lambda$. It reads
\begin{equation}
dX\,X^{-1}=\frac{\lambda}{\lambda^2+\rho^2}\left[-\lambda\,M^{-1}dM+\rho\ast M^{-1}dM\right]\,.
\end{equation}
Defining a generating matrix field $\Psi$ as
\begin{equation}
\Psi(\lambda,x)=M(x)\,X(\lambda,x)\qquad \textrm{with}\qquad \Psi(0,x)=M(x)\, ,
\end{equation}
we note that $\Psi$ satisfies
\begin{equation}
d\Psi\,\Psi^{-1}=\frac{\rho}{\lambda^2+\rho^2}\left[\rho\,dM+\lambda \star dM\right]M^{-1}\,. \label{eqV}
\end{equation}
$d\Psi$ in this equation can be calculated as follows
\begin{equation}
\begin{aligned}
d\Psi(\lambda,x)=&\,\,d\Psi(\lambda,x)\big|_\lambda+\frac{\partial \Psi(\lambda,x)}{\partial\lambda}\,d\lambda\\
=&\left[\partial_\rho \Psi+\frac{2\,\lambda\,\rho}{\lambda^2+\rho^2}\,\partial_\lambda \Psi\right]d\rho+
\left[\partial_z \Psi-\frac{2\,\lambda^2}{\lambda^2+\rho^2}\,\partial_\lambda \Psi\right]dz\\
=&\,\,(D_2\Psi)\,d\rho+(D_1\Psi)\,dz\,,
\end{aligned}
\end{equation}
where we have used \eqref{eqlambda} when going from the first line to the second one. We immediately see that $D_1$ and $D_2$ are precisely the differential operators introduced by BZ, defined in \eqref{defD1D2}. Therefore, the $\rho$ and $z$ components of equation \eqref{eqV} are given by
\begin{equation}
D_2\Psi=\frac{\rho\,U+\lambda\,V}{\lambda^2+\rho^2}\,\Psi\,,\qquad
D_1\Psi=\frac{\rho\,V-\lambda\,U}{\lambda^2+\rho^2}\,\Psi\,, \label{LaxM}
\end{equation}
where $U=\rho\,\partial_\rho M\,M^{-1}$ and $V=\rho\,\partial_z M\,M^{-1}$, which is precisely the BZ Lax pair (\ref{Laxpair}).

\subsection{Group theoretic structure and solution generation}
\label{grouptheory}

Although the Lax equations are equivalent in the two formalisms, the explicit constructions of  solutions from a given seed solution  of the higher dimensional theory are very different. In the rest of this section, we very briefly discuss the group theoretic structure construction of a new solution from a given seed solution in the formalism of \cite{Breitenlohner:1986um, Nicolai:1991tt}, to which we refer the reader for further details.

It is useful to start by recalling how the underlying symmetries appear in three dimensions, and how they can be used as a solution generating technique. The situation in two dimensions can then be seen as an infinite dimensional generalization of it.

The matrix $\cV$ constructed in three dimensions is an element of the coset $G_{2(2)}/\tilde K$, where the local isotropy group $\tilde K$ is $\tilde K=SL(2,\RR)\times SL(2,\RR)$. In our construction the matrix $\cV$ is taken to be in the Borel gauge (also known as the triangular gauge). The three-dimensional scalar Lagrangian \eqref{lagM} is invariant under global $G_{2(2)}$ transformations. The transformation of the matrix $\cV$ under $G_{2(2)}$ is not simply by the right multiplication of the group element $g$. This is because in general such a multiplication does not preserve the Borel gauge. In order to restore the Borel gauge a local transformation under an element of $\tilde K$ is needed:
\beq
\cV(x)\rightarrow k(x)\cV(x)g,
\eeq
where $g\in G_{2(2)}$ and $k(x)\in \tilde K$ for each $x$. The Lagrangian is  invariant under these transformations.  Finding the right compensator $k(x)$ is in general a very difficult problem. To get round this difficulty, one uses the defining property of elements $k\in\tilde K$,  namely $k^\sharp k=1$, where $k^\sharp:=\eta^{-1}k^{T}\eta$ (see equation (\ref{defsharp})). Thus, one can see that the matrix $\mathcal{M}=\cV^{\sharp}\cV$ transforms as $\mathcal{M}\rightarrow g^\sharp \mathcal{M} g.$ Similarly, the matrix $M$ defined in (\ref{M}) transforms as $M\rightarrow \tilde g^TM \tilde g$, where $\tilde g = S^{-1} g S$. The matrices $\mathcal{M}$ or $M$ easily allow us to construct new solutions starting from a seed solution.

Some of these considerations generalize to the theory obtained after reduction to two dimensions, where one now defines a matrix $\hat\cV(x,\gamma)$ that depends on a complex parameter $\gamma$ and is such that $\hat\cV(x,\gamma=0)=\cV(x)$. This matrix is therefore an element of the loop extension\footnote{The loop extension of a group $G$ is part of the untwisted affine extension of $G$. The extra elements in the latter correspond the central element and the derivation of the algebra, which can be seen to have an action on, respectively, the confomal factor $\xi$ and the  size $\rho$ of the internal space \cite{Bernard:1997et,Kleinschmidt:2005bq}. See for example \cite{Kac} or \cite{Fuchs} for more details on the theory of affine groups and algebras.} of $G_{2(2)}$, which we denote $G_{2(2)}^+$ and global transformations under elements of this infinite dimensional group should lead to new solutions of the same theory. A generalization of the triangular gauge is obtained by imposing that $\hat\cV(x,\gamma)$ is regular at $\gamma=0$, which is convenient as this limit defines $\cV(x)$. As in three dimensions, it is therefore necessary to perform a compensating transformation, this time under an element $\hat k(x,\gamma)$ such that $\hat k^\sharp\left(x,-\frac1\gamma\right)\hat k(x,\gamma)=\mathbf1$. This defines a subgroup of $G_{2(2)}^+$, which we note $\tilde K^{\infty}$. In summary, in two dimensions, the transformation of $\hat\cV(x,\gamma)$ is of the form
\beq
\hat\cV(x,\gamma)\rightarrow \hat k(x,\gamma)\hat\cV(x,\gamma)\hat g(w),
\eeq
where $\hat g(w)\in G_{2(2)}^+$ and $\hat k(x,\gamma)\in \tilde K^{\infty}$ for each $x$. To avoid having to find the right compensator $k(x,\gamma)$ and to work only with global transformations, one might want to try something similar to the case of three dimensions. Indeed, in the formalism of \cite{Breitenlohner:1986um, Nicolai:1991tt} one defines the monodromy matrix $\hat \cM$, which is,
\be
\hat \cM := \hat \cV^{\sharp}\left(x, -\frac{1}{\gamma}\right)\hat \cV(x,\gamma)~.
\label{Nsolgen}
\ee
A short calculation then reveals that $\hat \cM$ only depends on the spacetime independent spectral parameter $w$, i.e., it is independent of the spacetime coordinates $\rho, z$:
\be
\hat \cM  = \hat \cM (w).
\ee
The solution generating method then consists in multiplying the seed matrix $\hat \cM(w)$ with an appropriate group element $g(w)$ of the loop group $G_{2(2)}^+$ so that the new matrix is
\be
\hat \cM'(w) = g^{\sharp}(w) \hat \cM(w) g(w).
\ee
Now one needs to find  the factorization of the new  $\hat \cM'(w)$ as in \eqref{Nsolgen}.  Once the matrix $\hat \cV' (\gamma,x)$ is obtained, the matrix $ \cV' (x)$ can be readily obtained by taking the $\gamma\to0$ limit, and from there one can read all scalars. From the scalars one can reconstruct the higher dimensional solution. However, in order to be able to extract the new solution, the matrix $\hat \cV$ should be chosen in the triangular gauge, so that in \eqref{Nsolgen} the first factor is analytic in a neighborhood of $\gamma=0$, and the second factor is analytic in the neighborhood of $\gamma=\infty$. Moreover,  the overlap of the respective domains of analyticity should be an annulus region \cite{Breitenlohner:1986um, Nicolai:1991tt}. Finding such a factorization is a variant of the famous matrix valued Riemann Hilbert problem. Consequently, avoiding the need to find the compensator $\hat k(x)$ is done at the price of another equally difficult problem. However, when the matrix $\hat \cM(w)$ is chosen to be just consisting of simple poles in the complex $w$ plane, such a factorization can be reduced to an algebraic problem \cite{Nicolai:1991tt}. But even then, one problem remains: how to decide  which $\hat \cM(w)$ to start with, and  which $g(w)$ to multiply with, in order to generate an interesting solution of the higher dimensional theory?

In the approach of \cite{Breitenlohner:1986um, Nicolai:1991tt}, although the group theoretic aspects of the corresponding generalized Geroch group remain clear, from the practical stand point, the construction of new solutions has to-date remained obscure. For this reason, in the next section we generalize the BZ construction to the $G_{2(2)}/(SL(2,\RR) \times SL(2, \RR))$ coset model. The BZ method is also a purely algebraic procedure, but it has the advantage that it is more practical from the point of constructing new and interesting solutions of the higher dimensional theories.

A detailed dictionary between the BZ construction and the approach of \cite{Breitenlohner:1986um, Nicolai:1991tt} is still lacking. We will discuss this and related points further in section \ref{sec:open_problems}.

\section{Belinski-Zakharov Construction}
\label{sec:gen_BZ}
In section \ref{BZ} we generalize the BZ construction to minimal supergravity in terms of  the $7\times7$ matrix $M$ . In section  \ref{sec:factorspace} we discuss the action of the BZ transformations on the so called  factor space.

\subsection{The inverse scattering method for minimal supergravity in 5d}
\label{BZ}
Belinski and Zakharov provided a purely algebraic procedure for constructing  solutions to the Einstein vacuum equations  with $D-2$ commuting Killing fields.  Their method can be  adapted to the case of minimal supergravity. Since the procedure is essentially the same as in the vacuum case we shall be brief (see \cite{BZ} for more details).

The BZ Lax equations \eqref{Laxpair} are linear; therefore, we can construct a new solution to the Lax pair  by `dressing' a known `seed' solution $\Psi_0(\lambda,\rho,z)$. 
Defining a dressing matrix $\chi(\lambda, \rho, z)$, one seeks a new solution to \eqref{Laxpair}  of the form
\be
\Psi = \chi \Psi_0\,.\label{newpsi}
\ee
Making this substitution into equation \eqref{Laxpair} results in the equations for $\chi$.  The matrix $\chi$ is further constrained to ensure that the new matrix $M(\rho,z)=\Psi(\lambda=0,\rho,z)$ is real and symmetric.

We are interested in `solitonic' solutions which are characterized by having only simple poles in the complex $\lambda$-plane. From \eqref{newpsi}, it follows that for a general $n$-soliton transformation, the matrix $\chi$ will add $n$ new poles to the seed solution $\Psi_0$.  Furthermore, we shall restrict ourselves to cases where the poles are located on the real axis; this determines uniquely the location of the poles to be \cite{BZ}
\begin{equation}
 \tilde\mu_k=\pm\sqrt{\rho^2+(z-\omega_k)^2}-(z-\omega_k)\,,\label{muk}
\end{equation}
where the $\omega_k$ are real constants. The `$+$' and `$-$' poles are commonly known as solitons and antisolitons respectively.

A general $n$-soliton transformation, in addition to the $n$ real constants  $\omega_k$, is determined by $n$ arbitrary real seven-dimensional\footnote{For minimal supergravity we use a $7\times7$ representation of Lie algebra $\mf{g_{2(2)}}$. For other theories (or for other representations of $\mf{g_{2(2)}}$ for minimal supergravity) the dimension of these vectors depend on the dimension of the appropriate representation.} constant vectors $m_0^{(k)}$. These are known as the BZ vectors, and in the present context, their components control the addition of angular momentum and charges to a given seed solution.

Starting from a seed matrix $M_0$, an $n$-soliton transformation yields a new matrix $M$,
\begin{equation}
\label{finsol}
M_{ab}=(M_0)_{ab}-
  \sum_{k,l=1}^{n}\frac{
  (M_0)_{ac}\, m_c^{(k)}\,  (\Gamma^{-1})_{kl}\;  m_d^{(l)}\, (M_0)_{db}}
                       {\tilde\mu_k\tilde\mu_l}\; ,
\end{equation}
where the repeated indices are summed. The vectors $m_a^{(k)}$ are given by,
\begin{equation}\label{mvec}
m_a^{(k)}=m_{0b}^{(k)}\left[\Psi_0^{-1}(\tilde\mu_k,\rho,z)\right]_{ba}\,,
\end{equation}
and the matrix $\Gamma$ is defined as,
\begin{equation}\label{gammamat}
\Gamma_{kl}=\frac{m_a^{(k)}\, (M_0)_{ab}\, m_b^{(l)}}{\rho^2+\tilde\mu_k \tilde\mu_l}\; .
\end{equation}

The new matrix $M$, \eqref{finsol}, does not satisfy $\det
M=1$, and hence it is not a physically acceptable
solution\footnote{Since $G_{2(2)}$ is a subgroup of
$SO(3^+, 4^-)$, all $G_{2(2)}$ matrices have determinant equal to
unity.}. For cases when the matrix $M$ is $2 \times 2$, e.g., for
four dimensional vacuum gravity, one can deal with this problem by
multiplying it with a suitable factor of $\rho$ and $\tilde\mu_k$'s
\cite{BZ}. When the matrix $M$ is not $2 \times 2$ this rescaling
typically leads to nakedly singular solutions. A  clever and practical
solution to this problem has been proposed in \cite{Pomeransky:2005sj}
that can deal with the general case.
We follow the approach of
\cite{Pomeransky:2005sj} for our $7 \times 7$ matrix $M$.

The main idea is the observation that the determinant of the matrix $\eqref{finsol}$ does not depend on the BZ vectors.
Thus, one may start with a vacuum static solution $(M_0,\xi_0)$ (hence diagonal
matrix $M_0$) and remove some solitons or
antisolitons $\tilde\mu_k$ at $z=w_k$ with `trivial' vectors, say, with the ones aligned
with one of the basis vectors, that is,  $$m_{0b}^{(k)} =\delta_{ab} \qquad  \mbox{for
some given $a$}~.$$ This amounts to rescaling (cf.~\eqref{finsol})  the $aa$ component of the
matrix $M_0$ while leaving all other components unchanged:
\beq\label{remsoliton}
(\widetilde{M}_0)_{aa}=-\frac{\tilde\mu_k^2}{\rho^2}(M_0)_{aa}\,,\qquad
(\widetilde{M}_0)_{bb}=(M_0)_{bb} \qquad \mbox{with} \qquad b\neq a~.
\eeq
Next one re-adds  the same solitons and antisolitons to the matrix
$\widetilde{M}_0$ but now with more general vectors
$m_{0a}^{(k)}$.\footnote{Note that the freedom to rescale these
vectors by a constant can be used to set one component of each of
these vectors to unity.} Following this procedure, it is ensured by
construction that the final matrix $M$ and initial matrix $M_0$ have
the same determinant.

Finally, the two-dimensional conformal factor $\xi$ for the new
solution is obtained from the seed solution as
\begin{equation}\label{e2nu}
 \xi^2= \xi^2_0 \;\sqrt{\frac{\det\Gamma}{\det\Gamma_{0}}}\,,
\end{equation}
where the matrices $\Gamma_{0}$ and $\Gamma$ are obtained from
\eqref{gammamat} using $M_0$ and $M$ respectively. Compared to the
expression given for $e^\nu$ in \cite{Pomeransky:2005sj} for example,
we have an extra square root. This is due to the extra $1/2$ factor in
equations (\ref{lambdaeq}) for $\xi$  compared to equations
(\ref{nueq})  for $\nu$. Note that the extra term in the second set of
equations, compared to the first one, does not play a role in this
calculation (see also footnote \ref{footnotenu}).

There is, however, one serious problem. The new matrix $M$ in
\eqref{finsol}, obtained following the procedure described above, is
in general not an element of the coset $G_{2(2)}/(SL(2,\RR) \times
SL(2, \RR))$. A physical solution of minimal supergravity must be
represented by a matrix that is in the coset. We address this problem
in section \ref{stayinthecoset}.

\subsection{Action of the BZ transformations on the factor space}
\label{sec:factorspace}
In this subsection we discuss the action of the BZ transformations on the so-called factor space. For the class of solutions we are interested in, the factor space can be identified \cite{Hollands:2007aj} as the two-dimensional subspace orthogonal to the Killing vectors. We can use $\rho$ and $z$ as coordinates on this space \cite{Hollands:2007aj,Chrusciel:2008js}. These coordinates cover the region outside the horizon(s) and away from the rotational axes. The global structure of the factor space has been discussed in detail in \cite{Hollands:2007aj}, so we shall be brief. Considering spacetimes containing non-extremal horizons\footnote{See \cite{Figueras:2009ci} for the generalisation to the extremal case.}, the corresponding factor space can be shown to be a two-dimensional manifold possessing a connected boundary with corners. The boundary consists of a union of intervals. The intervals either correspond to the fixed points of the rotational Killing vectors or to  the horizon(s). The corners are the points where two adjacent intervals meet.

For vacuum five-dimensional gravity one can invariantly define a set of parameters on the factor space that uniquely characterize the solution \cite{Hollands:2007aj}. These parameters are the lengths of the intervals and certain vectors associated to the intervals. The vectors specify the linear combination of the Killing vectors that vanishes on the given interval.

 The presence of the gauge field in minimal supergravity generalizes this set of parameters \cite{Tomizawa:2009tb, Armas:2009dd, Hollands:2007qf}. The important point for this discussion is that the set of intervals together with the associated lengths and vectors is the same as in the vacuum case.  This is because they  only specify where the axes of symmetry and the horizons are in the spacetime. Based on this observation we now argue that in minimal supergravity the action of the BZ  method  on the factor space is very similar to the case of vacuum five-dimensional gravity.

To understand the action of the BZ  method on the factor space let us first consider four-dimensional vacuum gravity   reduced to two dimensions. The action of the BZ transformations on the seed generating matrix $\Psi_0$ consists of adding new poles in the complex $\lambda$-plane at the location of the solitons. Since any four dimensional black hole spacetime within this class can be obtained from Minkowski space, this implies that the action of the BZ transformations on the factor space consists of adding new intervals together with the corresponding vectors\footnote{Recall that for four dimensional Minkowski space, the boundary of the factor space is a single infinite interval.}. The constants added to the seed solution via the BZ vectors are related to the angular momentum and the NUT charges of the new solution.

In five-dimensional vacuum gravity the action of the BZ transformations can be understood in very much the same way. In practice the BZ method in five dimensions is implemented following the  approach of \cite{Pomeransky:2005sj}. This consists of starting with a static seed solution with (roughly speaking) the desired intervals, followed by removing solitons and re-adding them using BZ.  Unlike in four dimensions where one normally starts with Minkowski space, this procedure does not add new intervals to the boundary of the factor space. Removing solitons generically renders the seed solution singular; the Killing vectors no longer vanish in a smooth way on the axes (in fact some may diverge).  However, the factor space still has the same set of intervals as the seed solution. Re-adding the solitons adds new poles so as to cure the singularities that appeared when removing the solitons\footnote{The final spacetime may still contain singularities that may or may not be removable by a choice of parameters.}. Associated to each interval there will now be a non-trivial vector related to the direction of the solitons.

The above considerations are quite general, and do not {\em per se} refer to the matter fields of the theory. Therefore we expect that for minimal supergravity the action of the BZ dressing method on the factor space is essentially the same as it is for vacuum gravity. Of course, since in minimal supergravity the BZ vectors are seven dimensional, their components in general will be related to the angular momenta, NUT and the other charges of the final solution. We use this intuition in the practical implementation of the BZ method for minimal supergravity in the following sections.

\section{Some general results}
\label{sec:general_results}

In this section we establish some general results  of interest for the practical implementation of the BZ construction. In section \ref{subsec:commute} we show that the action of the three-dimensional hidden symmetry transformations
on the BZ dressing method is simply the group action on the BZ vectors. In section \ref{stayinthecoset} we discuss the issue of whether or not the BZ dressing method preserves the coset.

\subsection{Combining hidden symmetries and the BZ construction}
\label{subsec:commute}

In this section we  explicitly show that the action of the three-dimensional
hidden symmetry transformations on the BZ dressing method is simply the group
action on the BZ vectors.

\begin{figure}
\begin{center}
\includegraphics[width=7cm]{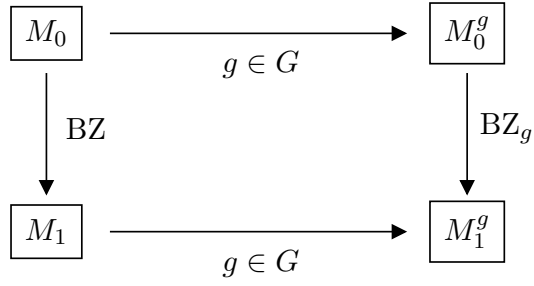}
\caption{This figure depicts the action of the hidden symmetry transformations on the BZ construction. Let the matrix $M_0$ be sent to $M_0^g$ by an element $g\in G$ and to $M_1$ by a BZ transformation. Let $M_1$ be sent to the matrix $M_1^g$ under the action of $g$. Then $M_1^g$  can also be obtained from  $M_0^g$ by applying the same BZ transformation but with BZ vectors appropriately transformed under $g$.}
\label{HScommutesBZ}
\end{center}
\end{figure}

  Consider the matrices $M_0$ and $M_0^g$ that are related by an element $g$ of the three-dimensional symmetry group\footnote{In our case, $G=G_{2(2)}$, but the results of this section apply to more general situations.} $G$  as follows\footnote{The precise transformation of the matrix $M$ under global $G_{2(2)}$ transformations is \eqref{g2transformation}. To avoid notational clutter we simply use $g$ to mean $S^{-1}gS$ in this section. }
\beq
M_0^g(x)=g^TM_0(x)g.
\eeq
Next consider the matrix $M_1$ obtained by applying a certain BZ transformation on $M_0$. We write this as
\beq
\label{BZtrans}
\Psi_1(\la,x)&=&\chi(\la,x)\widetilde\chi(\la,x)\Psi_0(\la,x)\nn\\
&\equiv&\chi(\la,x)\widetilde\Psi_0(\la,x).
\eeq
where $\Psi_0(\la,x)$ and $\Psi_1(\la,x)$ are the generating matrices corresponding respectively to $M_0(x)$ and $M_1(x)$, $\widetilde\chi(\la,x)$ is the dressing matrix that removes solitons from $\Psi_0(\la,x)$  and $\chi(\la,x)$ is the one that adds solitons to $\widetilde\Psi_0(\la,x)\equiv\widetilde\chi(\la,x)\Psi_0(\la,x)$. Applying the group transformation $g$ to the matrix $M_1$ leads to another matrix $M_1^g$:
\beq
M_1^g(x)=g^TM_1(x)g.
\eeq
We show that by performing the same BZ transformation  on $M_0^g$ (with appropriately modified BZ vectors to be discussed below)  one gets exactly $M_1^g$. Schematically, the situation is presented in figure \ref{HScommutesBZ}.

It is useful to first notice that the generating matrices $\Psi$ transform under group elements in the same way as the corresponding $M$ matrices. That is, the generating matrices $\Psi_{0,1}^g$ that solve the Lax equations~(\ref{Laxpair}) and satisfy $\Psi_{0,1}^g(\la=0,x) = M_{0,1}^g(x)$ are simply
\beq
\Psi_{0,1}^g(\la,x) = g^T \: \Psi_{0,1}(\la,x) \: g\, .
\label{RNfromSchw}
\eeq
Using this and equation (\ref{BZtrans}), and inserting $\mathbf1=(g^T)^{-1}g^T$ one has
\bea
\Psi_1^g(\la,x) &=& g^{T} \: \Psi_1(\la,x) \: g\nn\\
                      &=&  \left( g^{T} \: \chi(\la,x) \: (g^{T})^{-1} \right) \left( g^{T} \: \widetilde\Psi_0(\la,x) \: g \right)\label{interp1}\\
                        &\equiv&  \left( g^{T} \: \chi(\la,x) \: (g^{T})^{-1} \right)\widetilde\Psi_0^g~.\nn
\eea
This equation tells us that if one starts with the matrix obtained by removing solitons on $\Psi_0(\la,x)$ and transforming it with the element $g\in G$, then the matrix $\Psi_1^g$ is obtained by performing a BZ transformation using $\hat\chi(\la,x)\equiv g^{T} \: \chi(\la,x) \: (g^{T})^{-1}$ as the dressing matrix. In the following we study in more detail the new dressing matrix $\hat\chi$ and show that it adds the same solitons as $\chi$ but with BZ vectors transformed appropriately under $g$.

Let us now consider the dressing matrix $\hat\chi$. It follows from the BZ construction~\cite{BZ} that it obeys
\beq
\hat\chi =g^{T}\: \chi(\la,x) \: (g^{T})^{-1} = 1 + \sum_{k = 1}^{n} \frac{g^{T} \: R_k(x) \: (g^{T})^{-1}}{\lambda - \tilde \mu_k (x)}\, ,
\eeq
so we can concentrate on the residue matrices $\hat R_k(x) \equiv g^{T} \: R_k(x)\: (g^{T})^{-1} $. From the general BZ construction~\cite{BZ} we know that the matrices $R_{k}$ are degenerate and factorize as
\beq
(R_k)_{ab} = n^{(k)}_a m^{(k)}_{b}\, .
\eeq
Note that $a,b, \ldots$ are group representation indices. The $n^{(k)}$ are naturally column vectors whereas the $m^{(k)}$ take the form of row vectors. Thus,
\beq
\hat R_k = g^T \: R_k\: (g^T)^{-1}
         = \left( g^T n^{(k)} \right) \left( m^{(k)} (g^{-1})^T \right)
         \equiv  \hat n^{(k)} \hat m^{(k)}\, .
\label{Rkhat}
\eeq
A short calculation yields
\bea
\hat m^{(k)}  &\equiv&  m^{(k)} (g^{-1})^T
                =  m_0^{(k)} \widetilde{\Psi}_0^{-1}(\tilde \mu_k,x) (g^{-1})^T  \no \\
                &=&  \left( m_0^{(k)} g \right) ({\widetilde{\Psi}_{0}^g})^{-1}(\tilde \mu_k,x)
                \equiv \hat m_0^{(k)} ({\widetilde{\Psi}_{0}^g})^{-1}(\tilde \mu_k,x)\,,
\eea
where expressions~(\ref{mvec}) and the definition for $\widetilde{\Psi}_{0}^g$ have been employed in the second and third equalities, respectively.
Regarding the vector $\hat n^{(k)}$, we now show that its implicit definition from the factorization of the residue matrix~(\ref{Rkhat}) matches the BZ prescription for the construction of $M_1^g$ from $M_0^g$. The latter determines $\hat n^{(k)}$ in terms of $\hat m^{(k)}$ and $\widetilde{M}_0^g(x)\equiv\widetilde{\Psi}_{0}^g(\la=0,x)$. Indeed, recall that~\cite{BZ}
\beq
\hat n^{(k)}_{a} = \sum_{l} \tilde \mu^{-1}_{l} \hat D^{kl} \hat N_a^{(l)},\qquad \hat N_a^{(l)} = \hat m^{(l)}_{c} \widetilde{M}_0^g(x)_{ca}\,
\eeq
where
 $\hat D^{kl}$ is the inverse of the matrix $\hat \Gamma_{kl}$ introduced in \eqref{gammamat}. Next, observe that the matrix $\hat \Gamma_{kl}$ that takes us from $M_0^g$ to $M_1^g$ is exactly the same as the matrix $\Gamma_{kl}$ that takes us from $M_0$ to  $M_1$:
\beq
\hat \Gamma_{kl} &=& \frac{ \left( m^{(k)} (g^{-1})^T \right) \left( g^{T} \widetilde{M}_0(x) g \right)
                          \left( m^{(l)} (g^{-1})^T \right)^T }{\rho^2 + \tilde \mu_k \tilde \mu_l}\nonumber\\
                 &=& \frac{ m^{(k)} \widetilde{M}_0(x) (m^{(l)})^T }{\rho^2 + \tilde \mu_k \tilde \mu_l}\nonumber\\
                 &=& \Gamma_{kl}\, .
\eeq
It follows immediately that $\hat D^{kl}=D^{kl}$. Finally, consider the vector $\hat N^{(l)}$. It is also related to its untransformed counterpart in a simple manner:
\beq
(\hat N^{(l)})^T  &=&  \hat m^{(l)} \widetilde{M}_0^g(x)\nonumber\\
                  &=&  \left( m^{(l)} (g^{-1})^T \right) \left( g^T \widetilde{M}_0(x) g \right)\nonumber\\
                  &=&  m^{(l)} \widetilde{M}_0(x) g\nonumber\\
                  &=&  (N^{(l)})^T g\, .
\eeq
Therefore,
\beq
\hat n^{(k)} = \sum_{l} \tilde \mu^{-1}_{l} \hat D^{kl} \hat N^{(l)}
             = g^T \sum_{l} \tilde \mu^{-1}_{l} D^{kl} N^{(l)}
             = g^T n^{(k)}\, ,
\eeq
in accordance with~(\ref{Rkhat}).

In conclusion, we have established that to generate $M_1^g$ from $M_0^g$ using the inverse scattering method all we need to do is to make the replacement
\beq
m^{(k)}_{0} &\longrightarrow& \hat m^{(k)}_{0} = m^{(k)}_{0} g\,, \quad \mbox{and} \\
\widetilde{\Psi}_0(\la,x) &\longrightarrow& \widetilde{\Psi}_0^g(\la,x) = g^T\: \widetilde{\Psi}_0(\la,x)\: g\,,
\eeq
in the calculation that generates $M_1$ from $M_0$. Thus, the BZ construction of $M_1^g$ from $M_0^g$ and of  $M_1$ from $M_0$ are essentially the same.  In particular, this allows us to easily obtain the Cveti$\check{\rm c}$-Youm (CY) solution by applying the BZ construction on Reissner-Nordstr\"om (RN). The latter is obtained from Schwarzschild by a $G_{2(2)}$ transformation~\cite{Bouchareb:2007ax, G2}, so all we need to know is how to derive Myers-Perry (MP) from Schwarzschild using inverse scattering, and this is accomplished in section~\ref{subsec:Schw2MP}. In section \ref{subsec:RN2CY} we also explicitly obtain the Cveti$\check{\rm c}$-Youm (CY) solution from Reissner-Nordstr\"om (RN).

The result of this section supports the idea that the inverse
scattering method is
incapable of producing charging transformations since BZ transformations
are essentially insensitive to the presence of charges generated by
the 
three-dimensional hidden symmetry transformations.

\subsection{Do we stay in the coset after the BZ transformations?}
\label{stayinthecoset}

In this subsection we address the issue of whether or not the BZ transformations preserve the coset.
We start by discussing a toy model where similar problems arise but can be easily resolved. Then we discuss the case of minimal supergravity where the resolution seems to require to impose an extra non-linear constraint (equation \eqref{constraint} below) on the parameters of the solution.

Suppose for a moment that instead of minimal supergravity we were working with some other theory with the following coset model structure:
\be
\frac{SL(3,\RR)}{SO(2,1)} \times \frac{SL(3,\RR)}{SO(2,1)}~.
\ee
The higher dimensional origin of this coset model is not important for the point we want to make\footnote{Although a higher dimensional origin of the $\frac{SL(3,\RR)}{SO(2,1)} \times \frac{SL(3,\RR)}{SO(2,1)}$ coset model is not immediately obvious, appropriate truncation of heterotic theory does give rise to $\frac{SL(2,\RR)}{SO(2)} \times \frac{SL(2,\RR)}{SO(2)}$ coset model, see e.g., \cite{Bakas}. A similar structure also arises in certain truncations of five-dimensional Einstein-Maxwell theory \cite{Yazadjiev:2006hw}.}. The coset representative matrix $M$ for this theory will have the form
\be
M  = \left(
\begin{array}{cc}
M_{\rom{SL(3)}}^1 & 0  \\
0 &  M^2_{\rom{SL(3)}}
\end{array}
\right) \, ,
\ee
where $M^1_{\rom{SL(3)}}$ and $M^2_{\rom{SL(3)}}$ are two independent matrices representing two $\frac{SL(3,\RR)}{SO(2,1)}$ coset representatives. Multisoliton solutions of this theory will be characterized by two integers $(n_1,n_2)$, referring to the number of solitons associated to each $\frac{SL(3,\RR)}{SO(2,1)}$ coset model.  For each soliton in each $SL(3,\RR)$ there is a freedom to add three free parameters, namely, the BZ vector\footnote{Note that there is a  freedom to rescale each of these vectors by a constant and therefore, without loss of generality, we can always set one of the components of each vector to unity. For the sake of simplicity of the argument, we are not concerned with the normalization of the BZ vectors in this discussion.} associated to the soliton. Therefore, for $(n_1, n_2)$ solitons we can add a maximum of $3 (n_1 +  n_2) $ free parameters and still maintain the block diagonal form of the coset representative. This number is obviously less than $6 (n_1 + n_2)$, which is the number of free parameters one might have excepted very naively from the fact that the matrix $M$ is six dimensional. This theory has an additional symmetry, which we call ``double image'' symmetry: every solution possesses a companion solution that is obtained by interchanging the field content of the two $SL(3,\RR)/SO(2,1)$ coset models. Two solutions related by this symmetry may admit very different interpretations.

The situation in minimal supergravity is somewhat similar. Recall that the seven dimensional representation of $G_{2(2)}$ branches into $SL(3, \RR)$ representations as
\be
{\textbf 7} = \bar{\textbf{3}}  + \textbf{1} + \textbf{3}.
\ee
The $\textbf{3}$ and $\bar{\textbf{3}}$ represent two appearances of vacuum gravity in the matrix $M$:
\beq
M  = \left(
\begin{array}{ccc}
M_{\rom{SL(3)}}^{-1} & 0 & 0 \\
0  & 1 & 0 \\
0 & 0 & M_{\rom{SL(3)}}
\end{array}
\right) \, .
\label{vacuum1}
\eeq
Unlike the case discussed above,  $M_{\rom{SL(3)}}$ and $M_{\rom{SL(3)}}^{-1}$  both represent exactly the same $\frac{SL(3,R)}{SO(2,1)}$ coset model. We call this situation a ``mirror image''. Therefore, in order to maintain the form \eqref{vacuum1},  when we add $n$-solitons to the matrix $M_{\rom{SL(3)}}$ and $3 n$ free parameters through BZ vectors, we necessarily have to add $n$-antisolitons to the matrix $M_{\rom{SL(3)}}^{-1}$.  The BZ vectors associated to the antisolitons have to be related to the solitonic BZ vectors in such a way that the final matrix $M$ has precisely the block diagonal form \eqref{vacuum1}. It is intriguing that in all the examples we have looked at this can be achieved by relating the antisolitonic BZ vectors to the solitonic BZ vectors in a simple way. Though, a priori, it is not clear if it can be achieved for a general solution of vacuum gravity.

In the general case, the matrix $M$ representing a solution of minimal supergravity has the symmetrical block diagonal structure \eqref{cosetM} \cite{Bouchareb:2007ax, Clement2}
\beq
M = \left(
\begin{array}{ccc}
A & U & B \\
U^T & \tilde S & V^T \\
B^T & V & C
\end{array}
\right),
\eeq
where $A$ and $C$ are symmetric $3\times 3$ matrices, $B$ is a $3\times 3$ matrix, $U$ and $V$ are 3-component column matrices, and $\tilde S$ a scalar. The inverse matrix is given by \eqref{cosetMinv} \cite{Bouchareb:2007ax, Clement2}
\beq
M^{-1} = \left(
\begin{array}{ccc}
C & -V & B^T \\
- V^T & \tilde S &-U^T \\
B & -U & A
\end{array}
\right).
\eeq
We observe that
\begin{itemize}
\item Any solution of minimal supergravity should admit a limit to a solution  of vacuum gravity, and since vacuum gravity is encoded two times in the $7 \times 7$ matrix $M$, once as $M_{\rom{SL(3)}}$ and second time as $M_{\rom{SL(3)}}^{-1}$, it is natural to think that the field content of minimal supergravity is also `encoded two times' in the matrix $M$.
\item This observation is further strengthened by the fact that the inverse of the matrix $M$ is simply given by a reshuffling of the rows and columns (up to certain minus signs) of the matrix $M$.
\end{itemize}
Based on these observations, we expect that the soliton transformations on matrix $M$ in the general case are always accompanied with antisoliton transformations.

In practice, when we apply solitonic and antisolitonic transformations with general BZ vectors we move out of the coset. (We discuss an explicit example where this problem arises in the next section.)  This is a serious problem, because a physical solution of minimal supergravity must
be represented by a matrix that is in the coset.   Since the coset construction is based on the exceptional Lie group $G_{2(2)}$ one cannot write a simple matrix criterion for a $7\times7$ matrix to be in the coset. The precise conditions for a matrix $M$ obtained through BZ construction to be in the coset are the following
\begin{enumerate}
\item The matrix $\mathcal{M} := \cV^{\sharp} \cV = \eta^{-1} (S^T)^{-1}M S^{-1}$ should preserve the three form $c_{(\mf a)abc}$ constructed in appendix \ref{3form} in the sense that
\be
\sum_{a,b,c = 1}^{7}c_{(\mf{a})}{}_{abc} \mathcal{M}_{ae} \mathcal{M}_{bf} \mathcal{M}_{cg} = c_{(\mf a)}{}_{efg}~,
\label{constraint}
\ee
\item The  matrix $M$ should be symmetric.
\end{enumerate}
Roughly speaking, the first condition ensures that the final matrix $M$ is in $G_{2(2)}$ and the second condition, given the first condition, ensures that it is in the coset $G_{2(2)}/(SL(2, \RR) \times SL(2, \RR))$. The BZ construction guarantees the second condition, but it does not guarantee the first one. We believe that through certain relationships between solitonic and  antisolitonic BZ vectors one might ensure that the first condition is also satisfied.  However, how to find these precise conditions in the general case is not clear to us.  Therefore, at this stage, one should regard equation \eqref{constraint} as an extra non-linear constraint that every physical solution has to satisfy. Further exploration of these questions is left for the future.

\section{Examples}
\label{sec:examples}
As an illustration of our formalism, we obtain the doubly spinning five-dimensional Myers-Perry black hole by applying solitonic transformations on the Schwarzschild black hole (section \ref{subsec:Schw2MP}). We also derive the Cveti$\check{\rm c}$-Youm black hole by applying solitonic transformations on the Reissner-Nordstr\"om black hole (section \ref{subsec:RN2CY}).

\subsection{5d Myers-Perry from 5d Schwarzschild}
\label{subsec:Schw2MP}

The five-dimensional Myers-Perry solution has been constructed from 5d Schwarzschild-Tangherlini using the  BZ method in~\cite{Pomeransky:2005sj} in the context of vacuum gravity. However, we are now considering five-dimensional minimal supergravity and so this calculation provides a good testing ground for our framework.

For 5d Schwarzschild the metric  in the parametrization~(\ref{Killing+conformal}) takes the form
\beq \label{Schw}
&G_{\bar \mu \bar \nu}^\rom{Schw} = {\rm diag} \left\{G_{tt}^\rom{Schw},G_{\phi\phi}^\rom{Schw},G_{\psi\psi}^\rom{Schw} \right\}={\rm diag} \left\{ -\frac{\mu_1}{\mu_2}, \mu_2, \frac{\rho^2}{\mu_1} \right\}\,,&\nn \\
&\left(e^{2\nu}\right)_\rom{Schw} = \frac{\mu_2(\rho^2+\mu_1\mu_2)}{(\rho^2+\mu_1^2)(\rho^2+\mu_2^2)}\,.&
\eeq
In terms of the $SL(3, \RR)$ matrix of~\cite{Giusto:2007fx}, the Killing metric part translates into
\beq
M_{\rom{SL(3)}} = {\rm diag} \left\{ G_{tt}^\rom{Schw}, G_{\psi\psi}^\rom{Schw}, \frac{1}{G_{tt}^\rom{Schw} G_{\psi\psi}^\rom{Schw}} \right\}
       = {\rm diag} \left\{ -\frac{\mu_1}{\mu_2}, \mu_2, -\frac{1}{\mu_1} \right\}\,.\nonumber
\eeq
Since this is a vacuum solution we know from~(\ref{vacuum_mt}) that
\beq
M_\rom{Schw} = {\rm diag} \left\{ M_{\rom{SL(3)}}^{-1}, 1, M_{\rom{SL(3)}} \right\}
       = {\rm diag} \left\{ -\frac{\mu_2}{\mu_1}, \frac{1}{\mu_2}, -\mu_1, 1, -\frac{\mu_1}{\mu_2}, \mu_2, -\frac{1}{\mu_1} \right\}\,.\nonumber
\eeq
It is known that the five-dimensional doubly spinning Myers-Perry solution can be obtained from Schwarzschild-Tangherlini by a two-soliton transformation~\cite{Pomeransky:2005sj}. In the context of 5d minimal supergravity this, in practice, amounts to a four-soliton transformation on $M_\rom{Schw}$.  This is because each transformation on the $M_{\rom{SL(3)}}$ block  must  be accompanied by a corresponding transformation on the $M_{\rom{SL(3)}}^{-1}$ block.

To obtain the seed matrix we first remove an antisoliton at position $z=w_1$ with BZ vector $(\vec{0},0,\vec{e}_1)$ and, accordingly, remove a soliton at the same position but with BZ vector $(\vec{e}_1,0,\vec{0})$. Here we are defining $\vec{e}_1 \equiv (1,0,0)$ for convenience. The second pair consists of removing a soliton (respectively  an antisoliton) at position $z=w_2$ with the same BZ vectors as the former antisoliton (respectively a soliton). This whole procedure can be implemented by multiplying $M_\rom{Schw}$ by
\beq
H_\rom{Schw} = {\rm diag} \left\{ \frac{\mu_1^2}{\mu_2^2}, 0,0,0, \frac{\mu_2^2}{\mu_1^2}, 0,0 \right\}\,.
\eeq
In addition, we rescale the matrix thus obtained by $(\mu_1\mu_2)^{-1}$ to avoid the appearance of divergences during the BZ construction. Hence, our seed matrix is
\beq
\widetilde{M}_\rom{Schw}  \equiv  (\mu_1\mu_2)^{-1} H_\rom{Schw} M_\rom{Schw}
   = {\rm diag} \left\{ -\frac{1}{\mu_2^2}, \frac{1}{\mu_1\mu_2^2}, -\frac{1}{\mu_2}, \frac{1}{\mu_1\mu_2}, -\frac{1}{\mu_1^2}, \frac{1}{\mu_1}, -\f{1}{\mu_1^2\mu_2} \right\}\,,\nonumber
\eeq
and the associated solution of the Lax pair~(\ref{Laxpair}), $\widetilde{\Psi}_\rom{Schw}(\lambda,\rho,z)$, is obtained by shifting $\mu_k\rightarrow\mu_k-\lambda$, where $k=1,2$.

To perform the `dressing' we re-add the same (anti-)solitons, this time with more general BZ vectors:
\begin{itemize}
\item add an antisoliton at $w_1$ with $m^{(1)}_0 = \left( \vec{0},0,A_1,B_1,0 \right)$,
\item add a soliton at $w_1$ with $\bar{m}^{(1)}_0 = \left( \bar{A}_1,\bar{B}_1,0,0,\vec{0} \right)$,
\item add a soliton at $w_2$ with $m^{(2)}_0 = \left( \vec{0},0,A_2,0,C_2 \right)$,
\item add an antisoliton at $w_2$ with $\bar{m}^{(2)}_0 = \left( \bar{A}_2,0,\bar{C}_2,0,\vec{0} \right)$.
\end{itemize}
The new matrix, $M_{\rom{MP}}$, is then obtained from~(\ref{finsol}). It is now convenient to use the freedom of translating the $z$ coordinate so that the (anti-)solitons are placed at $w_2=\alpha=-w_1$. The mere fact that the first $3\times 3$ block of $M$ must be the inverse of the last $3\times 3$ block for the vacuum truncation imposes two restrictions on the BZ vectors:
\beq
\bar{C}_2 = 4\alpha C_2 \frac{\bar{A}_2}{A_2}\,, \qquad \bar{B}_1 = \frac{B_1}{4\alpha} \frac{\bar{A}_1}{A_1}\,. \label{conditions}
\eeq
If we do not impose these conditions then the matrix $M_{\rom{MP}}$ is not an element of the coset $G_{2(2)}/(SL(2, \RR) \times SL(2,\RR))$. The conditions~\eqref{conditions} are precisely the solution of the non-linear constraint (\ref{constraint}) in this example.
With these conditions the new matrix $M_{\rom{MP}}$ depends only on four parameters, which we take to be $\{A_1, B_1, A_2, C_2\}$. Moreover, if we take $B_1,C_2 \rightarrow 0$ we recover the static solution. This is in complete agreement with reference~\cite{Pomeransky:2005sj}. From the matrix $M_{\rom{MP}}$ we then extract the five scalars $\phi_1, \phi_2, \chi_1, \chi_5$ and $\chi_6$ that are turned on for this solution using~(\ref{chisl3}).

To construct the conformal factor for the dressed solution we first note that
\beq
e^{2\nu} = e^{\frac{1}{\sqrt{3}}\phi_1+\phi_2} \xi^2\,.
\eeq
from~(\ref{metric5}) and~(\ref{gmunu}).
Applying the transformation~(\ref{e2nu}) to the conformal factor for the Schwarzschild solution given in~(\ref{Schw}) we obtain the conformal factor for the dressed solution. To compare it with the Myers-Perry one we relate the components of the BZ vectors to the angular momentum parameters as
\be
B_1=l_1 A_1 \sqrt{\frac{4\al}{\beta}}\, , \qquad
C_2=\frac{l_2 A_2}{\sqrt{4\al\beta}}\, ,
\ee
where we have defined $\beta \equiv 2\al + \sqrt{l_1^2 l_2^2 +4\al^2}\,$ for convenience.
As mentioned in section~\ref{inverse}, there is still a multiplicative constant of integration
which is undetermined by the equations of motion for the conformal factor but it can be fixed by the requirement of asymptotic flatness. Expressed in prolate spherical coordinates (see appendix~\ref{CY}) the conformal factor for Myers-Perry becomes
\beq
\left(e^{2\nu}\right)_\rom{MP} = \frac{4\al x + (l_1^2 - l_2^2)y + 2m}{8\al^2 (x^2 - y^2)}\,,
\eeq
where
$
2m \equiv l_1^2 + l_2^2 + 2\sqrt{l_1^2 l_2^2 +4\al^2}.
$

To recover the full five-dimensional metric one then needs to follow the inverse route of the procedure outlined in section~\ref{3dreduction}. Namely, one dualizes the new scalars $\chi_5$ and $\chi_6$ back into one-forms and then reconstructs the metric using~(\ref{metric5}). In order to identify the new solution as Myers-Perry, one needs to perform a linear transformation mixing the coordinates $\{t,\phi,\psi\}$. The latter is easily determined by imposing that the angular coordinates match asymptotically with those for flat spacetime. This transformation reads
\bea
t &\longrightarrow & t - \frac{l_1\left( l_1^2-l_2^2 -4\al +\beta \right)}
   {l_1^2-l_2^2}\,\phi+ \frac{l_2^3}{l_1^2-l_2^2}\, \psi\,, \nn \\
\phi &\longrightarrow & - \frac{l_1 l_2}{\sqrt{4\al\beta}}\,\phi + \sqrt{\frac{\beta}{4\al}}\,\psi\,, \nn \\
\psi &\longrightarrow & \sqrt{\frac{\beta}{4\al}}\,\phi - \frac{l_1 l_2}{\sqrt{4\al\beta}}\,\psi\,.
\label{coc}
\eea
The metric thus obtained is the Myers-Perry solution expressed in prolate spherical coordinates. This is displayed in appendix~\ref{CY}, with the understanding that the uncharged solution has $\delta=0$.

\subsection{5d Cveti$\check{\rm c}$-Youm solution from 5d Reissner-Nordstr\"om}
\label{subsec:RN2CY}
Using the derivation of the previous subsection and the results of section \ref{subsec:commute} we know how to construct the Cveti$\check{\rm c}$-Youm solution by applying a solitonic transformation on the Reissner-Nordstr\"om solution. In this subsection, we present the main steps of this calculation.

One first needs to construct the matrix $M$ for the Reissner-Nordstr\"om solution. This is most easily done as \cite{G2}
\beq
M_\rom{RN}=g^TM_\rom{Schw}g
\eeq
for $g=S^{-1}\, e^{\sqrt3 k_2 \delta }\,S$
where the matrix $S$ and the generator $k_2$ are defined in appendices \ref{genG2} and \ref{rep} while $\delta$ is the charge parameter appearing in the Reissner-Nordstr\"om solution (given in appendix \ref{CY} if one takes $l_1=l_2=0$). The corresponding generating matrix is similarly obtained as
\beq
\Psi_\rom{RN}=g^T\Psi_\rom{Schw}g.
\eeq
As discussed in section \ref{subsec:commute}, the seed solution is obtained by removing from $\Psi_\rom{RN}$ the same solitons and antisolitons as in the vacuum case but with BZ vectors that are now non-trivial: they are equal to $(\vec{0},0,\vec{e}_1)g$ and $(\vec{e}_1,0,\vec{0})g$. This is equivalent to taking the seed to be
\beq
\widetilde\Psi_\rom{RN}=g^T\widetilde\Psi_\rom{Schw}g\, \qquad  \mathrm{and} \qquad \widetilde M_\rom{RN}=g^T\widetilde M_\rom{Schw}g\, .
\eeq

To perform the dressing we add the same solitons and antisolitons as in the vacuum case but with BZ vectors taken to be
\beq
\widehat m^{(1)}_0=m^{(1)}_0 g\, ,\qquad \widehat {\bar m}^{(1)}_0=\bar{m}^{(1)}_0 g,\\
 \widehat m^{(2)}_0=m^{(2)}_0 g\,,\qquad \widehat {\bar m}^{(2)}_0 =\bar{m}^{(2)}_0 g.
 \eeq
In this case, we already know all five conditions to impose on  the eight BZ parameters.
It is reassuring that the new matrix $M_\rom{CY}$ is in the coset and that the new conformal factor matches with the Cveti$\check{\rm c}$-Youm solution.

From $M_\rom{\rom{CY}}$ one can reconstruct
the five-dimensional metric and gauge field. As for the Myers-Perry case, the identification of the solution requires a linear change of coordinates:
\beq
t &\longrightarrow & t +c_1\,\phi+ c_2\, \psi\,, \nn \\
\phi &\longrightarrow & - \frac{l_1 l_2}{\sqrt{4\al\beta}}\,\phi + \sqrt{\frac{\beta}{4\al}}\,\psi\,, \nn \\
\psi &\longrightarrow & \sqrt{\frac{\beta}{4\al}}\,\phi - \frac{l_1 l_2}{\sqrt{4\al\beta}}\,\psi\,.
\eeq
where
\beq
c_1&=&- \frac{l_1\left( l_1^2-l_2^2 -4\al +\beta \right)c^3-l_2\left( -l_1^2+l_2^2 -4\al +\beta \right)s^3}
   {(l_1^2-l_2^2)}\,,\nn\\
c_2 &=&\frac{l_2^3\,c^3-l_1^3\,s^3}{(l_1^2-l_2^2)}\, ,\nn
\eeq
and $s = \sinh \delta$ and $c = \cosh \delta$.
It is easy to see that this change of coordinates simplifies to (\ref{coc}) when $\delta=0$.
The metric and the gauge field thus obtained are the ones for Cveti$\check{\rm c}$-Youm expressed in prolate coordinates, which is presented in section \ref{CY}.

\section{Ehlers versus Matzner-Misner reduction}
\label{EvsMM}

In this section we argue that to see the integrability of five-dimensional minimal supergravity one necessarily has to
perform an Ehlers reduction to two dimensions.

In sections \ref{sec:dimred} and \ref{inverse} we described two ways of reducing a theory to two dimensions. The first one, performed in section \ref{sec:dimred} for minimal five-dimensional supergravity, is in two steps. First one reduces to three dimensions and dualizes vectors into scalars. Second, one finally reduces to two dimensions. This two-step way of doing the reduction to two dimensions is called an Ehlers reduction. The second method, described in section \ref{inverse} for vacuum gravity in five dimensions, consists in reducing directly to two dimensions, without dualizations.   This way of doing the reduction is called a Matzner-Misner reduction.

The dualization of vectors into scalars in three dimensions can make some symmetries transparent. These are the so-called hidden symmetries.  For a series of theories \cite{Cremmer:1999du}, the hidden symmetry group ($G$) is larger  than (or equal to) the symmetry group obtained via dimensional reduction without dualization.  In all these cases the dimensionally reduced supergravity Lagrangian can be written in three dimensions, after dualization, as gravity plus a non-linear sigma model with the symmetry group $G$. Upon further reduction to two dimensions this hidden symmetry group $G$ gets enlarged to its untwisted affine extension, generally denoted as $G^{(1)}$. In all known examples, this untwisted affine extension of the group $G$ is the symmetry group of the dimensionally reduced theory in two dimensions. One way to show this is through the Lax pair in the Ehlers reduction, as explained in section \ref{grouptheory}. Thus, the untwisted affine extension of the group $G$ naturally leads to an integrable two-dimensional sigma model based on the group $G$. However, in some cases, the Matzner-Misner reduction also allows us to see this symmetry, and thus the integrability of the theory. This is the case, for example, for vacuum gravity. This is most easily understood by considering the problem from a slightly different perspective, namely in terms of Dynkin diagrams.

\begin{figure}[h!]
\begin{center}
\includegraphics[width=4cm]{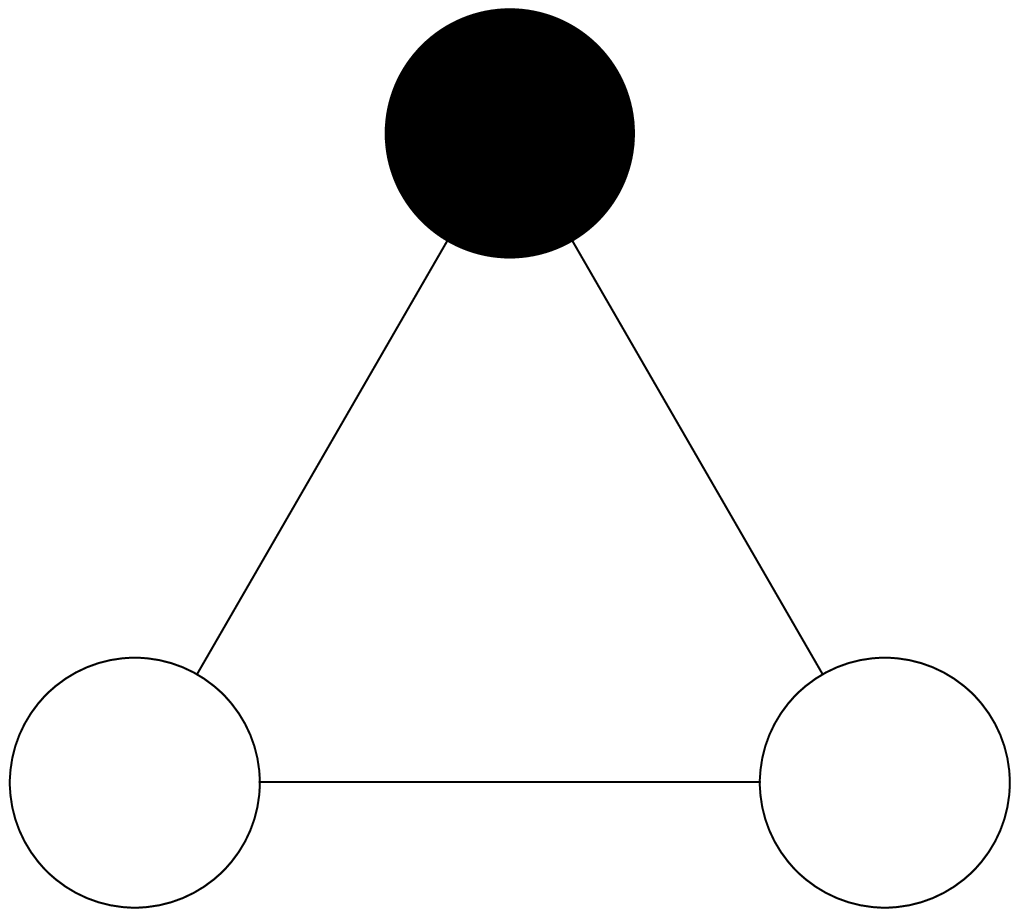}
\caption{Dynkin diagram of the untwisted affine extension of $\mf{sl(3,\RR)}$.}
\label{fig:dynkinsl3plus}
\end{center}
\end{figure}

Schematically one can also view the affine extension of $G$ appearing in the dimensional reduction to two dimensions in the following way (without invoking the Lax pair):
\begin{itemize}
\item Perform an Ehlers reduction to two dimensions: one then obtains a system of scalars $\Phi$ that transform non-linearly under $G$, which is the symmetry group in three dimensions.
\item Perform a Matzner-Misner reduction to two dimensions: one obtains another system of scalars $\bar \Phi$, related to $\Phi$ by non-local field redefinitions involving dualizations, and transforming in a certain way under $\bar G$.
\item Studying the intertwining of $G$ with $\bar G$, that is, studying how $G$ acts on $\bar \Phi$ and $\bar G$ acts on $\Phi$, one finds that they form an infinite-dimensional group which is the affine extension of $G$. This is how, for example, the infinite dimensional Geroch group \cite{Geroch:1972yt} of four dimensional general relativity was discovered (although the affine structure of the group was not clear at the time of \cite{Geroch:1972yt}).
\end{itemize}
In the case of five-dimensional vacuum gravity, the Ehlers group $G$ and the Matzner-Misner group $\bar G$ are both $SL(3, \RR)$. These two $SL(3, \RR)$ intertwine to give the affine group $SL(3, \RR)^+$, which is the symmetry of vacuum five-dimensional gravity when reduced to two dimensions. The Dynkin diagram of $SL(3, \RR)^+$ is shown in figure \ref{fig:dynkinsl3plus}. On the Dynkin diagram the Ehlers $SL(3, \RR)$ corresponds to the two white dots, while the Matzner-Misner $SL(3, \RR)$ is made out of the black dot and one of the white dots. 
In this case, as $G$ and $\bar G$ are the same groups, taking the Lax pair based on any of the two leads to the same infinite dimensional symmetry. The integrability can accordingly be seen from both point of views. This is the reason why we are able to perform a BZ construction based on the Killing sector of the metric (Matzner-Misner) as well as based on the matrix $M_{\rom{SL(3)}}$ (Ehlers).

\begin{figure}[h!]
\begin{center}
\includegraphics[width=6cm]{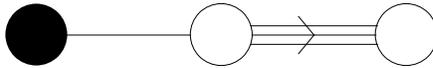}
\caption{Dynkin diagram of the untwisted affine extension of $\mf{g_{2(2)}}$.}
\label{fig:dynking2plus}
\end{center}
\end{figure}

For five-dimensional minimal supergravity, the Ehlers group $G$ is $G_{2(2)}$.  On the other hand, the Matzner-Misner group is $SL(3, \RR)$. Through a Matzner-Misner reduction, the scalars coming from the five-dimensional metric will form an $SL(3, \RR)/SO(2,1)$ coset model. These scalars will be coupled in a non-linear way with the degrees of freedom coming from the reduction of the five-dimensional gauge field $A^5_{(1)}$. On the Dynkin diagram of $G_{2(2)}^+$ given in figure \ref{fig:dynking2plus}, the Ehlers group $G_{2(2)}$ is given by the two white dots, while the Matzner-Misner $SL(3, \RR)$ is made of the black dot and the white dot connected to it. 
In this case, as $\bar G$ is not the same as $G$ but only a subgroup, taking the Lax pair based on $\bar G$ will not encompass the whole symmetry of the theory. (It is not even clear if such a Lax pair construction can be carried out.)  As a consequence, it is necessary to do the Ehlers reduction in order to see the integrability of five-dimensional supergravity in the presence of three commuting Killing vectors. In principle, with appropriate non-local field redefinitions one should be able to combine the $SL(3, \RR)$ fields in the Matzner-Misner reduction with the degrees of freedom of the dimensionally reduced gauge field $A^5_{(1)}$ and see an integrable sigma model based on the affine extension of $\mf{g_{2(2)}}$. We however believe that such field redefinitions would precisely correspond to an Ehlers reduction. Furthermore, we note that all theories based on classical Lie algebras, namely the $A_r, B_r, C_r$ and $D_r$ series of Lie algebras, give rise to the same integrable sigma model structure upon doing the reduction either the Ehlers way or the Matzner-Misner way. In particular, it is true for vacuum gravity in any dimension and for heterotic string theory \cite{Sen:1995qk}.

\section{Discussion and open problems}
\label{sec:open_problems}

One aim of this paper was to  explore the  integrability  of five-dimensional minimal supergravity in the presence of three
commuting Killing vectors with potential applications in mind.  Although it is well known in the supergravity literature \cite{Breitenlohner:1986um,  Nicolai:1991tt, Nicolai:1996pd} that the dimensionally reduced theory is completely integrable, the integrability structure has not been used as a solution generating technique.
In this paper we took steps in this direction. We derived the BZ Lax pair for minimal supergravity, using a $7\times 7$ symmetric coset representative matrix $M$, which is well suited for many applications.  We also generalized the well-known BZ construction; however, we are unable to provide precise conditions (to be imposed perhaps on the BZ vectors) that will ensure by construction that the dressed matrix is also  in the coset. Since the coset construction for this theory is based on the exceptional Lie group $G_{2(2)}$, checking whether a final matrix is in the coset is itself a non-trivial task. We provided an algebraic way for checking if the final matrix $M$ is in the coset.   Our method is  based on the invariant three-form for our representation of the Lie algebra $\mf{g_{2(2)}}$. An explicit construction of this three-form is given in appendix \ref{3form}.

In section \ref{equivalence}, following \cite{Breitenlohner:1986um} we clarified  the relationship between the BZ Lax pair and the Lax pair of \cite{Nicolai:1991tt, Breitenlohner:1986um}. However, the precise relationship between the BZ inverse-scattering construction and the  group theoretic solution generating approach  of \cite{Nicolai:1991tt, Breitenlohner:1986um} is far from clear. We believe that understanding the precise relationship between these two techniques will also give some insights into the problem of ensuring in the  BZ dressing method that the final matrix $M$ is in the coset by construction. A detailed exploration of these ideas is left for the future.

As an illustration of the usefulness of our formalism, we obtained the doubly spinning five-dimensional Myers-Perry black hole using the $7\times7$ matrix $M$ as a four soliton transformation on the (four solitons removed) Schwarzschild black hole. We also derived the Cveti$\check{\rm c}$-Youm black hole solution as a four-soliton transformation on the four-soliton removed Reissner-Nordstr\"om solution. The second calculation was achieved by first  studying in detail the action of the three-dimensional hidden symmetry transformations on the BZ construction.

In section \ref{EvsMM} we argued that to see the integrability of five-dimensional minimal supergravity one necessarily has to perform an Ehlers reduction. A direct dimensional reduction to two dimensions does not allow us to see the integrability of the theory in an easy way. This situation is in contrast with the situation in vacuum gravity \cite{BZ} and with many other theories, e.g., heterotic theory \cite{Sen:1995qk}, where the integrability can be seen either way.

 The next step would be to understand the physics of dipole charges in our formalism. This is especially interesting because the known dipole black ring \cite{Emparan:2004wy},  having only one non-zero angular momentum parameter, is not the most general dipole ring. A dipole black ring with two independent angular momenta is expected to exist.  Our formalism provides a framework where a construction of doubly spinning dipole black ring can in principle be carried out, unlike in the approach of \cite{Yazadjiev:2006hw} where such a construction is not possible in principle. In order to construct the doubly spinning dipole ring, one could first try to understand the dipole charge in our coset formalism and then add rotation on the two-sphere using solitonic transformations. Once this is achieved adding electric charge to this solution would be relatively easy; i.e., it can be done, say, using the charging transformations of \cite{Bouchareb:2007ax, G2}.  This line of investigation may very well let us discover the conjectured five parameter family of non-supersymmetric black rings in minimal supergravity.

\subsection*{Acknowledgements}
We thank Joan Camps, Sophie de Buyl, Roberto Emparan, Marc Henneaux, Axel Kleinschmidt, Josef Lindman H\"ornlund, James Lucietti, Jnanadeva Maharana, Hermann Nicolai, Jakob Palmkvist, and Simon Ross for interesting discussions. We are particularly grateful to Jakob Palmkvist for his patient explanations of octonions. PF is supported by an STFC rolling grant.  EJ was a FRS-FNRS bursar. EJ and AV were supported by IISN - Belgium (conventions 4.4511.06 and 4.4514.08) and by the Belgian Federal Science Policy Office through the Interuniversity Attraction Pole P6/11. JVR acknowledges financial support from {\it Funda\c{c}\~ao para a Ci\^encia e Tecnologia} (FCT)-Portugal through fellowship SFRH/BPD/47332/2008. PF and JVR would like to thank the IISN and the Theoretical Physics group at ULB for hospitality during the final stages of this work.

\appendix

\section{Cveti$\check{\rm c}$-Youm solution in canonical coordinates}
\label{CY}

A four parameter family of charged rotating black holes in five-dimensional minimal supergravity was obtained by Cveti$\check{\rm c}$ and Youm in~\cite{Cvetic:1996xz} by applying boosts and string dualities to the (neutral, rotating) Myers-Perry black hole in five dimensions.
The four parameters specifying the solution can be chosen to be the mass $M$, two angular momenta $J_{\phi,\psi}$,  and electric charge $Q_E$.
It is convenient to present the solution in terms of the quantities $m, l_1, l_2,$ and $\delta$ related to the above through~\cite{Cvetic:1997uw}
\bea
M       &=& \left( \frac{\pi}{4 G_5}\right) 3 m \left(1+2 s^2 \right) \ , \\
J_{\phi} &=& \left( \frac{\pi}{2 G_5}\right)m \left(l_1 c^3- l_2 s^3 \right) \ , \\
J_{\psi} &=& \left( \frac{\pi}{2 G_5}\right)m \left(l_2 c^3- l_1 s^3\right) \ , \\
Q_E  &=& \frac{1}{4 \pi} \int_{S^3_{\infty}} \left( \star F - \frac{F \wedge A}{\sqrt{3}}\right) = - 2 \sqrt{3} \pi  m c s  \ .
\eea
where for notational convenience we use $s := \sinh \delta$, and $c = \cosh \delta$.

The spacetime has three commuting Killing directions which are denoted by $t$, $\phi$ and $\psi$, the first coordinate being time-like whereas the last two represent angular directions.
The remaining coordinates parametrize the radial direction and another angle, respectively $r$ and $\tht$.
In these coordinates the metric and the gauge field read as follows
\be
ds^2 = g_{tt}\, dt^2 + 2g_{t\phi}\, dtd\phi + 2g_{t\psi}\, dtd\psi
     + g_{\phi\phi}\, d\phi^2 + g_{\psi\psi}\, d\psi^2 + 2g_{\phi\psi}\, d\phi\psi
     + g_{rr}\, dr^2 + g_{\tht\tht}\, d\tht^2  \ ,
\label{metricCY}
\ee
and the gauge field as
\be
A = A_{t} dt + A_{\phi} d \phi + A_{\psi} d\psi \ ,
\ee
with
\bea
g_{tt}       &=&  -\, \frac{\Sg(\Sg - 2m)}{(\Sg + 2ms^2)^2}  \ , \nn  \\
g_{t\phi}    &=&  -\, \frac{2m \sin^2\tht \left[ \Sg \left\{ l_1c^3 - l_2s^3 \right\} +2ml_2s^3\right] }{(\Sg + 2ms^2)^2}  \ , \nn \\
g_{t\psi}    &=&  -\, \frac{2m \cos^2\tht \left[ \Sg \left\{ l_2c^3 - l_1s^3 \right\} +2ml_1s^3\right] }{(\Sg + 2ms^2)^2}  \ , \nn
\\
g_{\phi\phi} &=&  \frac{\sin^2\tht}{(\Sg + 2ms^2)^2} \left[ (r^2+2ms^2+l_1^2)(\Sg + 2ms^2)^2 \right. \nn \\
  &+& \left. 2m\sin^2\tht \left\{ \Sg(l_1^2c^2-l_2^2s^2)
  + 4ml_1 l_2 c^3s^3 - 2ms^4(l_1^2c^2+l_2^2s^2) - 4m l_2^2 s^4 \right\} \right]  \ , \nn \\
g_{\psi\psi} &=&  \frac{\cos^2\tht}{(\Sg + 2ms^2)^2} \left[ (r^2+2ms^2+l_2^2)(\Sg + 2ms^2)^2 \right. \nn \\
  &+& \left. 2m\cos^2\tht \left\{ \Sg(l_2^2c^2-l_1^2s^2)
  + 4ml_1 l_2 c^3s^3 - 2ms^4(l_2^2c^2+l_1^2s^2) - 4m l_1^2 s^4 \right\} \right]  \ , \nn \\
g_{\phi\psi} &=&  \frac{2m \cos^2\tht \sin^2\tht \left[ l_1 l_2 \left\{ \Sg-6ms^4 \right\} + 2m(l_1^2+l_2^2)s^3c^3
                - 4ml_1 l_2 s^6 \right] }{(\Sg + 2ms^2)^2}  \ , \nn
\\ 
g_{rr}       &=&  \frac{r^2(\Sg + 2ms^2)}{(r^2+l_1^2)(r^2+l_2^2)-2mr^2}  \ , \nn \\
g_{\tht\tht} &=&  \Sg + 2ms^2  \ , \nn \\
A_{t} &=& \frac{2 \sqrt{3} m s c}{(\Sg + 2ms^2)} \nn \ , \\
A_{\phi} &=& - A_t  (l_1 c - l_2 s) \sin^2 \theta \nn \ , \\
A_{\psi} &=& - A_t  (l_2 c - l_1 s)\cos^2 \theta  \ .
\eea
For convenience we have defined
\be
\Sg(r,\tht)  \equiv  r^2 + l_1^2 \cos^2\tht + l_2^2 \sin^2\tht   \ .
\ee
Setting $\delta=0$ reproduces the five-dimensional MP black hole.

To make contact with the inverse scattering method we need to express the metric~(\ref{metricCY}) in canonical form,
\be
ds^2 =  G_{\bar \mu \bar \nu}\, dx^{\bar \mu} dx^{\bar \nu} + e^{2\nu} \left( d\rho^2 + dz^2 \right) \ , \qquad {\rm with} \; \rho = \sqrt{|\det G|} \ , \label{metricCanonical}
\ee
where the three-dimensional Killing part of the metric, $G_{\bar \mu \bar \nu}$, and $\nu$ depend only on the coordinates $(\rho, z)$.
It turns out that the determinant of the Killing part of~(\ref{metricCY}) is {\em exactly} the same as for the five-dimensional MP solution, namely
\be
\det g^{(3)} = - \left[ (r^2+l_1^2)(r^2+l_2^2) - 2mr^2 \right] \sin^2\tht \cos^2\tht \ .
\ee
Thus, we immediately obtain
\be
\rho = \frac{r \sin 2\tht}{2} \sqrt{r^2 \left(1+\frac{l_1^2}{r^2}\right) \left(1+\frac{l_2^2}{r^2}\right) - 2m} \ .
\ee
To determine $z=z(r,\tht)$ one just assumes separability and requires~(\ref{metricCY}) to take the form~(\ref{metricCanonical}).
Once again this yields the same expression as for 5-dimensional MP:
\be
z = \left( \frac{r^2}{2} - \frac{2m-l_1^2-l_2^2}{4} \right) \cos 2\tht  \ .
\ee

The conformal factor $\nu(r,\tht)$ is given by
\be
e^{2\nu} = \frac{ \Sg + 2ms^2 }{ \left[ (r^2+l_1^2)(r^2+l_2^2) - 2mr^2 \right] \cos^2 2\tht + \frac{1}{4} \left( 2r^2 + l_1^2 + l_2^2 - 2m \right)^2 \sin^2 2\tht }  \ .
\ee
Of course, we need to express $\nu$ in terms of the canonical coordinates $\{ \rho, z\}$.
To this end, it is useful to introduce the prolate spherical coordinates, $x \in [1,\infty)$ and $y \in [-1,1]$, defined by~\cite{Harmark:2004rm}
\be
x = \frac{2r^2+l_1^2+l_2^2-2m}{4\al}  \qquad  ,  \qquad  y = \cos 2\tht  \ ,
\ee
where
\be
\al = \frac{1}{4} \sqrt{(2m-l_1^2-l_2^2)^2 - 4l_1^2 l_2^2}  \ .
\ee
In terms of these coordinates the conformal factor simplifies,
\be
e^{2\nu} = \frac{ 4\al x + (l_1^2 - l_2^2)y + 2m(1+2s^2) }{ 8 \al^2 (x^2 - y^2) }  \ .
\ee
The last term, proportional to $s^2$, is the only difference relative to the five-dimensional MP case.

The components of the metric $G_{\bar \mu \bar \nu}$ and the gauge field $A_{\bar \mu}$ may be easily expressed in terms of the prolate spherical coordinates, as well.
One finds
\bea
G_{tt}       &=&  -\, \frac{ \left(4\al x + (l_1^2 - l_2^2)y\right)^2 - 4m^2 }{ \left\{ 4\al x + (l_1^2 - l_2^2)y + 2m(1+2s^2) \right\}^2 }  \ , \nn \\
G_{t\phi}    &=&  -\, \frac{2m (1-y) \left[ \left\{ 4\al x + (l_1^2 - l_2^2)y + 2m \right\} l_1c^3 - \left\{ 4\al x + (l_1^2 - l_2^2)y - 2m \right\} l_2s^3 \right] }{ \left\{ 4\al x + (l_1^2 - l_2^2)y + 2m(1+2s^2) \right\}^2 }  \ , \nn \\
G_{t\psi}    &=&  -\, \frac{2m (1+y) \left[ \left\{ 4\al x + (l_1^2 - l_2^2)y + 2m \right\} l_2c^3 - \left\{ 4\al x + (l_1^2 - l_2^2)y - 2m \right\} l_1s^3 \right] }{ \left\{ 4\al x + (l_1^2 - l_2^2)y + 2m(1+2s^2) \right\}^2 }  \ , \nn \\
G_{\phi\phi} &=&  \frac{(1-y)\left\{ 4\al x + l_1^2 - l_2^2 + 2m(1+2s^2) \right\} }{4}  \nn \\
  &+&  \frac{m(1-y)^2}{\left\{ 4\al x + (l_1^2 - l_2^2)y + 2m(1+2s^2) \right\}^2}
     \left[ \left\{ 4\al x + (l_1^2 - l_2^2)y + 2m \right\} \left( l_1^2c^2-l_2^2s^2 \right) \right. \nn \\
  && \qquad \left. + 8ml_1 l_2c^3s^3 - 4ms^4\left( l_1^2c^2+l_2^2s^2 \right) - 8ml_2^2s^4 \right]  \ , \nn
\eea
\bea
G_{\psi\psi} &=&  \frac{(1+y)\left\{ 4\al x - l_1^2 + l_2^2 + 2m(1+2s^2) \right\} }{4}  \nn \\
  &+&  \frac{m(1+y)^2}{\left\{ 4\al x + (l_1^2 - l_2^2)y + 2m(1+2s^2) \right\}^2}
     \left[ \left\{ 4\al x + (l_1^2 - l_2^2)y + 2m \right\} \left( l_2^2c^2-l_1^2s^2 \right) \right. \nn \\
  && \qquad \left. + 8ml_1 l_2c^3s^3 - 4ms^4\left( l_2^2c^2+l_1^2s^2 \right) - 8ml_1^2s^4 \right]  \ , \nn \\
G_{\phi\psi} &=&  \frac{2m (1-y^2) \left[ \frac{1}{2}l_1 l_2 \left\{ 4\al x + (l_1^2 - l_2^2)y + 2m(1-6s^4) \right\} + 2m(l_1^2+l_2^2)s^3c^3 - 4ml_1 l_2s^6 \right] }{ \left\{ 4\al x + (l_1^2 - l_2^2)y + 2m(1+2s^2) \right\}^2 }  \ , \nn \\
A_{t} &=& \frac{4 \sqrt{3} m cs}{\left\{4\al x + (l_1^2 - l_2^2)y + 2m(1+2s^2) \right\}} \ , \nn \\
A_{\phi} &=& - A_t (l_1 c - l_2 s)\frac{ (1-y) }{2} \ , \nn \\
A_{\psi} &=& - A_t (l_2 c - l_1 s) \frac{ (1+y) }{2}\ . \nn
\eea

The above components may be translated into canonical coordinates by using the following relations~\cite{Harmark:2004rm}:
\bea
\rho &=& \al \sqrt{(x^2-1)(1-y^2)}  \ , \qquad \qquad  z = \al x y \ ,   \\
x    &=& \frac{R_+ + R_-}{2\al}  \ , \qquad \qquad  y = \frac{R_+ - R_-}{2\al} \ ,  \\ &&{\rm with} \; \; R_\pm = \sqrt{\rho^2+(z\pm\al)^2} \ .\nn
\eea
This completes the task of expressing the CY metric~(\ref{metricCY}) in canonical form~(\ref{metricCanonical}).

\section{Generalities on $G_2$}
\label{genG2}

For the purpose of reference, here we collect some basic facts about the Lie algebra $\mf{g_{2}}$ and its split real form $\mf{g_{2(2)}}$.
For further details we refer the reader to standard references, such as \cite{Humphreys:1980dw}. See also \cite{G2}.

\begin{figure}[h!]
\begin{center}
\includegraphics[width=4cm]{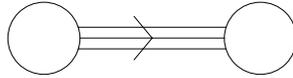}
\caption{Dynkin diagram of $G_2$.}
\label{fig:dynking2}
\end{center}
\end{figure}

The algebra $\mf{g_2}$ is the smallest of the exceptional Lie algebras.
 It has rank $2$ and its dimension is $14$.
Its Dynkin diagram is presented in figure~\ref{fig:dynking2}.
\noindent Each node of this diagram corresponds to a triple of Chevalley generators $\{H_a, E_a, F_a\}, a=1,2$.
The $H_a$'s span the Cartan subalgebra $\mf{h}$ of $\mf{g_2}$.
The $E_a$'s are the generators associated to the two simple roots $\vec\alpha_1$ and $\vec\alpha_2$ of $\mf{g_2}$. These generators satisfy the Chevalley relations
\bea
\left[ H_1, E_1\right] &=& 2 E_1~, \qquad \left[H_2, E_1\right] = - 3 E_1~, \qquad  \left[E_1, F_1\right] = H_1~, \nn \\
\left[H_1, E_2\right] &=& -E_2, \qquad \left[H_2, E_2\right] = 2 E_2~, \qquad \: \: \: \, \left[E_2, F_2\right] = H_2~.
\eea
The simple roots belong to the dual $\mf{h^\star}$ of $\mf{h}$.
By taking multiple commutators of $E_a$'s, and using Serre relations, one obtains a set of four more positive generators $E_k,\, k = 3, \ldots, 6$.
More explicitly, one can take them to be
\beq
E_3 = [E_1,E_2]\, ,  \hspace{1cm} E_4 = [E_3,E_2]\, ,  \hspace{1cm} E_5 = [E_4,E_2]\, ,  \hspace{1cm} E_6 =
[E_1, E_5]\,  . \label{genepos}
\eeq
\noindent In the basis
\bea
\label{newbasis}
h_1 &=& \frac{1}{\sqrt{3}} H_2, \quad  h_2 = H_2 + 2 H_1, \nn \\
e_1 &=& E_1, \quad e_2 = \frac{1}{\sqrt{3}}  E_2, \quad e_3 = \frac{1}{\sqrt{3}} E_3,\nn \\
e_4 &=& \frac{1}{\sqrt{12}} E_4, \quad  e_5 = \frac{1}{6} E_5, \quad e_6 = \frac{1}{6} E_6, \nn \\
f_1 &=& F_1, \quad f_2 = \frac{1}{\sqrt{3}}  F_2, \quad f_3 = \frac{1}{\sqrt{3}} F_3, \nn \\
f_4 &=& \frac{1}{\sqrt{12}} F_4, \quad  f_5 = \frac{1}{6} F_5, \quad  f_6 = \frac{1}{6} F_6.
\eea
the positive roots take the following values:
\begin{center}
\begin{tabular}{ll}
$\vec\alpha_1 = (-\sqrt{3},1)$,& \quad
$\vec\alpha_2 = (\frac{2}{\sqrt 3},0)$,\\
$\vec\alpha_3 = (-\frac{1}{\sqrt 3},1)=\vec\alpha_1+\vec\alpha_2$,& \quad
$\vec\alpha_4 = (\frac{1}{\sqrt 3},1)=\vec\alpha_1+2\vec\alpha_2$, \\
$\vec\alpha_5 = (\sqrt 3,1)=\vec\alpha_1+3\vec\alpha_2$,& \quad
$\vec\alpha_6 = (0,2)=2\vec\alpha_1+3\vec\alpha_2$.
\end{tabular}
\label{posroots}
\end{center}

We are interested in compactifying five-dimensional minimal supergravity over one timelike and one spacelike Killing direction. When one first compactifies along a direction of signature $\epsilon_1$ and then along a direction of signature $\epsilon_2$---where $\epsilon_{1,2}$ take values $+1$ or $-1$ depending upon whether the reduction is performed over a spacelike or a timelike direction---, the relevant involution $\tau$ is given as (for a more detailed discussion see \cite{G2}):
\begin{eqnarray}
\tau(h_1) &= &- h_1, \qquad  \tau(h_2)=-h_2, \no \\
  \tau(e_1) &= & -\eps_1 \eps_2 f_1, \qquad  \tau(e_2) = -\eps_1 f_2,\qquad   \tau(e_3) =  -\eps_2 f_3, \no \\
  \tau(e_4) &=&  -\eps_1 \eps_2 f_4,\qquad  \tau(e_5) =  -\eps_2 f_5, \qquad  \tau(e_6) = -\eps_1 f_6 \, . \label{inv}
\end{eqnarray}

\noindent The subalgebra of elements fixed under $\tau$ is not compact. It consists of all the elements of the form $\{e_i + \tau (e_i)\}$, that is,
\beq
k_1 &=& e_1 + f_1\, , \qquad k_2 \, = \, e_2+f_2 \, , \qquad k_3  = e_3-f_3 \, , \no \\
k_4 &=& e_4 + f_4 \, , \qquad k_5 \, =\,  e_5 - f_5\, , \qquad k_6 = e_6 + f_6  \, .\label{eq:k}
\eeq
These elements generate the $\mf{sl}(2,\RR)\oplus\mf{sl}(2,\RR)$ algebra.

\section{Representation of $\mf{g_{2(2)}}$ and coset representative}
\label{rep}

In this appendix we give a representation of $\mf{g_{2(2)}}$ and a construction of symmetric coset representative $M$ for the coset $G_{2(2)}/(SL(2, \RR) \times SL(2,\RR))$.
The coset construction given below is largely based on the one used in \cite{G2}, but it differs in one important aspect, namely that the final matrix $M$ given below is symmetric --- which is not the case in \cite{G2}.
In \cite{G2} the matrix $M$ is symmetric under generalized transposition, but not under the usual transpose.
The symmetric matrix given below is better suited for inverse scattering constructions.
The representation of $\mf{g_{2(2)}}$ we use is identical to the one used in \cite{G2}.
For completeness and for ease of reference here we present relevant details.
We start by defining the Chevalley generators:
\bea
E_1 =
\left(
\begin{array}{lllllll}
 0 & 0 & 0 & 0 & 0 & 0 & 0 \\
 0 & 0 & 1 & 0 & 0 & 0 & 0 \\
 0 & 0 & 0 & 0 & 0 & 0 & 0 \\
 0 & 0 & 0 & 0 & 0 & 0 & 0 \\
 0 & 0 & 0 & 0 & 0 & 1 & 0 \\
 0 & 0 & 0 & 0 & 0 & 0 & 0 \\
 0 & 0 & 0 & 0 & 0 & 0 & 0
\end{array}
\right),
&
&
E_2 = \left(
\begin{array}{lllllll}
 0 & 1 & 0 & 0 & 0 & 0 & 0 \\
 0 & 0 & 0 & 0 & 0 & 0 & 0 \\
 0 & 0 & 0 & 2 & 0 & 0 & 0 \\
 0 & 0 & 0 & 0 & 2 & 0 & 0 \\
 0 & 0 & 0 & 0 & 0 & 0 & 0 \\
 0 & 0 & 0 & 0 & 0 & 0 & 1 \\
 0 & 0 & 0 & 0 & 0 & 0 & 0
\end{array}
\right), \nn
\eea
\bea
F_1 = \left(
\begin{array}{lllllll}
 0 & 0 & 0 & 0 & 0 & 0 & 0 \\
 0 & 0 & 0 & 0 & 0 & 0 & 0 \\
 0 & 1 & 0 & 0 & 0 & 0 & 0 \\
 0 & 0 & 0 & 0 & 0 & 0 & 0 \\
 0 & 0 & 0 & 0 & 0 & 0 & 0 \\
 0 & 0 & 0 & 0 & 1 & 0 & 0 \\
 0 & 0 & 0 & 0 & 0 & 0 & 0
\end{array}
\right),
&
&
F_2 = \left(
\begin{array}{lllllll}
 0 & 0 & 0 & 0 & 0 & 0 & 0 \\
 1 & 0 & 0 & 0 & 0 & 0 & 0 \\
 0 & 0 & 0 & 0 & 0 & 0 & 0 \\
 0 & 0 & 1 & 0 & 0 & 0 & 0 \\
 0 & 0 & 0 & 1 & 0 & 0 & 0 \\
 0 & 0 & 0 & 0 & 0 & 0 & 0 \\
 0 & 0 & 0 & 0 & 0 & 1 & 0
\end{array}
\right), \nn
\eea
\bea
H_1 = \left(
\begin{array}{lllllll}
 0 & 0 & 0 & 0 & 0 & 0 & 0 \\
 0 & 1 & 0 & 0 & 0 & 0 & 0 \\
 0 & 0 & -1 & 0 & 0 & 0 & 0 \\
 0 & 0 & 0 & 0 & 0 & 0 & 0 \\
 0 & 0 & 0 & 0 & 1 & 0 & 0 \\
 0 & 0 & 0 & 0 & 0 & -1 & 0 \\
 0 & 0 & 0 & 0 & 0 & 0 & 0
\end{array}
\right),
&
&
H_2 =
\left(
\begin{array}{lllllll}
 1 & 0 & 0 & 0 & 0 & 0 & 0 \\
 0 & -1 & 0 & 0 & 0 & 0 & 0 \\
 0 & 0 & 2 & 0 & 0 & 0 & 0 \\
 0 & 0 & 0 & 0 & 0 & 0 & 0 \\
 0 & 0 & 0 & 0 & -2 & 0 & 0 \\
 0 & 0 & 0 & 0 & 0 & 1 & 0 \\
 0 & 0 & 0 & 0 & 0 & 0 & -1
\end{array}
\right)~.\nn
\eea
Using these generators we can easily write the representation matrices for the rest of the generators. Matrices in the basis (\ref{newbasis}) can then be readily obtained.

We write a coset representative $\cV$ for the coset $G_{2(2)}/(SL(2, \RR) \times SL(2,\RR))$ in the Borel gauge
by exponentiating the Cartan and positive root generators of $\mf{g_{2(2)}}$ with the dilatons and axions as coefficients.
We can make contact with the reduced Lagrangian (\ref{Lcoset_implicit}) by choosing the coset representative to be \cite{G2}
\begin{equation}
\mathbb \cV = e^{\half \phi_1 h_1 + \half \phi_2 h_2} e^{\chi_1 e_1 }e^{-\chi_2 e_2 +\chi_3 e_3}e^{\chi_6 e_6} e^{\chi_4 e_4 -\chi_5 e_5}.
\end{equation}
Next we introduce the following two matrices
\be
\eta = \left(
\begin{array}{lllllll}
 -\frac{1}{2} & 0 & 0 & 0 & 0 & 0 & 0 \\
 0 & \frac{1}{2} & 0 & 0 & 0 & 0 & 0 \\
 0 & 0 & -\frac{1}{2} & 0 & 0 & 0 & 0 \\
 0 & 0 & 0 & 1 & 0 & 0 & 0 \\
 0 & 0 & 0 & 0 & -2 & 0 & 0 \\
 0 & 0 & 0 & 0 & 0 & 2 & 0 \\
 0 & 0 & 0 & 0 & 0 & 0 & -2
\end{array}
\right),  \quad
S= \left(
\begin{array}{lllllll}
 0 & 0 & 0 & 0 & 0 & 0 & \sqrt{2} \\
 0 & -\sqrt{2} & 0 & 0 & 0 & 0 & 0 \\
 \sqrt{2} & 0 & 0 & 0 & 0 & 0 & 0 \\
 0 & 0 & 0 & -1 & 0 & 0 & 0 \\
 0 & 0 & 0 & 0 & \frac{1}{\sqrt{2}} & 0 & 0 \\
 0 & 0 & 0 & 0 & 0 & \frac{1}{\sqrt{2}} & 0 \\
 0 & 0 & \frac{1}{\sqrt{2}} & 0 & 0 & 0 & 0
\end{array}
\right),
\ee
and define the coset representative matrix $M$ as
\be
M = S^{T} \cV^T \eta \cV S~.
\ee
The matrix $M$ is symmetric by construction.
It can be easily checked by an explicit calculation that the scalar part of the reduced Lagrangian (\ref{Lcoset_implicit}) is given by
\be
\cL_{\rom{scalar}} = - \frac{1}{8} \mbox{Tr}\left(\star (M^{-1}d M) \wedge (M^{-1} d M)\right).
\ee
The matrix $M$ has the symmetrical block structure \cite{Bouchareb:2007ax, Clement2}
\beq
M = \left(
\begin{array}{ccc}
A & U & B \\
U^T & \tilde S & V^T \\
B^T & V & C
\end{array}
\right),\label{cosetM}
\eeq
where $A$ and $C$ are symmetrical $3\times 3$ matrices, $B$ is a $3\times 3$ matrix, $U$ and $V$ are 3-component column matrices, and $\tilde S$ a scalar.
It also follows that the inverse matrix is given by \cite{Bouchareb:2007ax, Clement2}
\beq
M^{-1} = (S^T m S) M (S^T m S) = \left(
\begin{array}{ccc}
C & -V & B^T \\
- V^T & \tilde S &-U^T \\
B & -U & A
\end{array}
\right), \label{cosetMinv}
\eeq
where $m$ is the $SO(3^+,4^-)$ metric \eqref{metric} for our representation.
Explicit expressions for the matrices $A,B,C,U,V,\tilde S$ can be readily written.
Since these expressions are exceedingly long, we do not present them here.
However, we note that
\bea
\det A = \det C.
\eea

\section[Invariant 3-form and the basis of octonions for our representation]{Invariant 3-form and the basis of octonions for our representation\footnote{This appendix is done in collaboration with  Josef Lindman H\"ornlund and Jakob Palmkvist.}}
\label{3form}

Recall that one definition of the complex Lie algebra $\mf{g_{2}}$ is as the algebra of endomorphisms of a seven dimensional vector space $V$ preserving a general three-form on $V$.
In this appendix we provide a construction of this invariant three-form $c_{(\mf a)}{}_{abc}$ for our representation\footnote{The reason why we choose the notation $c_{(\mf a)}{}_{abc}$ for the invariant three-form will become clear in the following.}.
This invariant three-form provides a way to check if a $7 \times 7$ matrix $\mathcal{M}$ is in the group $G_{2(2)}$ in our representation.
If a matrix $\mathcal{M}_{ab}$ preserves this three-form, i.e., if it satisfies
\be
\sum_{a,b,c = 1}^{7}c_{({\mf {a}})}{}_{abc} \mathcal{M}_{ae} \mathcal{M}_{bf} \mathcal{M}_{cg} = c_{(\mf a)}{}_{efg}
\ee
then it is in the group $G_{2(2)}$.

There are several ways to construct this three form.
A systematic way of doing this for a given representation of $\mf{g_{2}}$ is presented in e.g., \cite{FultonHarris}.
However, in this section we do not follow the approach of \cite{FultonHarris}; instead, we provide the octonion description of our representation.
The octonion structure constants $\tilde c_{(\mf a)}{}_{abc}$ are directly related to the invariant 3-form $c_{(\mf a)}{}_{abc}$.
The octonion description has the advantage of being more intuitive; and it also naturally allows us to introduce the metric $m_{ab}$ on the seven dimensional vector space $V$.

We start by recalling some basic facts about octonions.
The octonions form a real eight dimensional division algebra $\mathbf{O}$ with a basis $\{\mf e_0, \mf e_1, \mf e_2, \ldots, \mf e_7\}$ where $\mf e_0$ spans the real numbers.
The seven imaginary units $\{\mf e_1, \mf e_2, \ldots, \mf e_7\}$ anti-commute and square to $-1$.
When $a \neq b$, the product $\mf e_a \mf e_b$ is given in a standard basis, e.g., in the basis given in \cite{Baez}, by
\bea
\mf e_1 &=& \mf e_2 \mf e_4 = \mf e_3\mf e_7 = \mf e_5\mf e_6~, \nn \\
\mf e_2 &=& \mf e_3\mf e_5 = \mf e_4\mf e_1 = \mf e_6\mf e_7~, \nn \\
\mf e_3 &=& \mf e_4\mf e_6 = \mf e_5\mf e_2 = \mf e_7\mf e_1~,  \nn \\
\mf e_4 &=& \mf e_5\mf e_7 = \mf e_6\mf e_3 = \mf e_1\mf e_2~, \nn \\
\mf e_5 &=& \mf e_6\mf e_1 = \mf e_7\mf e_4 = \mf e_2\mf e_3~, \nn \\
\mf e_6 &=& \mf e_7\mf e_2 = \mf e_1\mf e_5 = \mf e_3\mf e_4~,\nn \\
\mf e_7 &=& \mf e_1\mf e_3 = \mf e_2\mf e_6 = \mf e_4\mf e_5~. \label{octbasis}
\eea
The complex Lie algebra $\mf{g_{2}}$ is the derivation algebra of the octonions $\mathbf{O}$.
A basis of the derivation algebra of $\mathbf{O}$ is given as
\bea
\mf e_{2L}\mf e_{4L} - \mf e_{5L}\mf e_{6L}~, &\qquad& \mf e_{2L}\mf e_{4L} - \mf e_{3L}\mf e_{7L}~, \nn \\
\mf e_{4L}\mf e_{1L} - \mf e_{3L}\mf e_{5L}~, &\qquad& \mf e_{4L}\mf e_{1L}-  \mf e_{6L}\mf e_{7L}~, \nn \\
\mf e_{1L}\mf e_{2L} - \mf e_{6L}\mf e_{3L}~, &\qquad& \mf e_{1L}\mf e_{2L} - \mf e_{5L}\mf e_{7L}~, \nn \\
\mf e_{4L}\mf e_{6L} - \mf e_{5L}\mf e_{2L}~, &\qquad& \mf e_{4L}\mf e_{6L} - \mf e_{7L}\mf e_{1L}~, \nn \\
\mf e_{1L}\mf e_{5L} - \mf e_{3L}\mf e_{4L}~, &\qquad& \mf e_{1L}\mf e_{5L}-  \mf e_{7L}\mf e_{2L}~, \nn \\
\mf e_{2L}\mf e_{3L} - \mf e_{6L}\mf e_{1L}~, &\qquad& \mf e_{2L}\mf e_{3L} - \mf e_{7L}\mf e_{4L}~, \nn \\
\mf e_{1L}\mf e_{3L} - \mf e_{2L}\mf e_{6L}~, &\qquad& \mf e_{1L}\mf e_{3L} - \mf e_{4L}\mf e_{5L}~,  \label{derivations}
\eea
where for any octonion $\mf e$ we define $\mf e_L$ as the endomorphism of $\mathbf{O}$ which acts on the left as
\be
\mf e_L: \mf o \rightarrow \mf e \mf o \qquad \mbox{for any} \ \mf o \in \mathbf{O}.
\ee
The derivation algebra is associative, unlike the octonions algebra $\mathbf{O}$.
To obtain the Chevalley generators of $\mf{g_{2}}$,  one takes the following linear combinations of the derivation elements \eqref{derivations}
\bea
E_1 &=& \frac{1}{4} \left[i(\mf e_{2L}\mf e_{3L} - \mf e_{7L}\mf e_{4L})  - \mf e_{7L}\mf e_{2L} + \mf e_{3L}\mf e_{4L}\right]~, \nn \\
F_1 &=& \frac{1}{4}
\left[i(\mf e_{2L}\mf e_{3L} - \mf e_{7L}\mf e_{4L}) + \mf e_{7L}\mf e_{2L} - \mf e_{3L}\mf e_{4L}\right]~, \nn \\
E_2 &=& \frac{1}{4} \left[i\left(2\mf e_{7L}\mf e_{1L} - \mf e_{4L}\mf e_{6L} - \mf e_{5L}\mf e_{2L}\right) - 2\mf e_{1L}\mf e_{3L} + \mf e_{2L}\mf e_{6L} + \mf e_{4L}\mf e_{5L}\right]~, \nn \\
F_2 &=& \frac{1}{4} \left[i\left(2\mf e_{7L}\mf e_{1L} - \mf e_{4L}\mf e_{6L} - \mf e_{5L}\mf e_{2L}\right) + 2\mf e_{1L}\mf e_{3L} - \mf e_{2L}\mf e_{6L} - \mf e_{4L}\mf e_{5L}\right]~, \nn \\
H_1 &=& \frac{1}{2} i\left(\mf e_{2L}\mf e_{4L} - \mf e_{3L}\mf e_{7L}\right)~, \nn \\
H_2 &=& \frac{1}{2} i \left(2\mf e_{3L}\mf e_{7L}- \mf e_{2L}\mf e_{4L} - \mf e_{5L}\mf e_{6L}\right)~.  \label{generators}
\eea
Since all elements of the derivation algebra act trivially on the real numbers, we consider only the action of
the elements \eqref{derivations} on the seven dimensional subspace $\mathbf{Im} \, \,  \mathbf{O}$ spanned by the imaginary units
$\{\mf e_1, \mf e_2, \ldots, \mf e_7\}$.
This provides the unique (up to a change of basis) seven dimensional representation of $\mf{g_{2}}$, the smallest
non-trivial representation.

In the basis
\be
\{\mf a_0,\mf a_1, \mf a_2, \mf a_3, \mf a_4, \mf a_5, \mf a_6, \mf a_7\}
\ee
where
\bea
\mf a_0 &=& \mf e_0 \nn \\
\mf a_1 &=& i\mf e_5 + \mf e_6 \nn \\
\mf a_2 &=& \mf e_2 + i \mf e_4 \nn \\
\mf a_3 &=& i\mf e_3 - \mf e_7 \nn \\
\mf a_4 &=& -2 i \mf e_1 \nn \\
\mf a_5 & = &-2 i \mf e_3 - 2 \mf e_7 \nn \\
\mf a_6 &=& -2 \mf e_2 + 2 i \mf e_4 \nn \\
\mf a_7 &=& -2 i \mf e_5 + 2 \mf e_6 \label{definition}
\eea
the $7\times 7$ matrices obtained from \eqref{generators} are precisely the ones given in appendix \ref{rep}\footnote{Note the use of $i$'s in the definitions \eqref{definition}. This is because we are working with the split real form of $\mf{g_2}: \mf{g_{2(2)}}$. Alternatively, one can use a basis of split octonions instead of \eqref{octbasis}. Then the entries in the corresponding matrix $A$, \eqref{matrixA}, would be all real.}.
An easy way to see this is to write the matrix $A$ that takes us from one basis to another.
Columns of the matrix $A$ are the $\mf a$-basis vectors expressed in the $\mf e$-basis,
\be
A = \left(
\begin{array}{lllllll}
 0 & 0 & 0 & -2 i & 0 & 0 & 0 \\
 0 & 1 & 0 & 0 & 0 & -2 & 0 \\
 0 & 0 & i & 0 & -2 i & 0 & 0 \\
 0 & i & 0 & 0 & 0 & 2 i & 0 \\
 i & 0 & 0 & 0 & 0 & 0 & -2 i \\
 1 & 0 & 0 & 0 & 0 & 0 & 2 \\
 0 & 0 & -1 & 0 & -2 & 0 & 0
\end{array}
\right)~. \label{matrixA}
\ee
Matrices in our representation of $\mf{g_{2(2)}}$ are precisely
\be
A^{-1} T_i A
\ee
where $T_i$ are the matrices obtained from \eqref{generators}.
The octonion structure constants in the $\mf e$-basis
\be
\mf e_a \mf e_b =  \sum_{c=1}^{7}\tilde c_{(\mf e)}{}_{abc}\mf e_{c}
\ee
and in the $\mf a$-basis
\be
\mf a_a \mf a_b = \sum_{c=1}^{7}\tilde  c_{(\mf a)}{}_{abc}\mf a_{c}.
\ee
are related by the basis change
\be
\tilde c_{(\mf a)}{}_{mnp} = \sum_{a,b,c = 1}^{7}A_{am} A_{bn} \tilde c_{(\mf e)}{}_{abc} (A^{-1})_{pc}~.
\ee
To construct the invariant three form we multiply the structure constants with the $SO(3^+,4^-)$ metric
\be
c_{(\mf a)}{}_{abc} = \sum_{e = 1}^{7}m_{ce}\tilde c_{(\mf a)}{}_{abe}~. \label{lower}
\ee
The $SO(3^+,4^-)$ metric is obtained using the matrix $A$
\be
m = A^T A =
\left(
\begin{array}{lllllll}
 0 & 0 & 0 & 0 & 0 & 0 & 1 \\
 0 & 0 & 0 & 0 & 0 & -1 & 0 \\
 0 & 0 & 0 & 0 & 1 & 0 & 0 \\
 0 & 0 & 0 & -1 & 0 & 0 & 0 \\
 0 & 0 & 1 & 0 & 0 & 0 & 0 \\
 0 & -1 & 0 & 0 & 0 & 0 & 0 \\
 1 & 0 & 0 & 0 & 0 & 0 & 0
\end{array}
\right). \label{metric}
\ee


\end{document}